\documentstyle[pre,aps]{revtex}

\def\Kappa{\mbox{\ae}}

\begin{document}
\draft
\title{Conformational transitions of heteropolymers in dilute solutions}
\author{E.G.~Timoshenko\thanks{Corresponding author. 
Internet: http://fiachra.ucd.ie/\~{}timosh},
Yu.A.~Kuznetsov, K.A.~Dawson}
\address{Theory and Computation Group,
Irish Centre for Colloid Science and Biomaterials\thanks{
Established at the University College Dublin and Queen's University of Belfast 
},\\
Department of Chemistry, University College Dublin,
Belfield, Dublin 4, Ireland}
\date{\today}
\maketitle

\begin{abstract}
In this paper we extend the Gaussian self--consistent method
to permit study of the equilibrium
and kinetics of conformational transitions for heteropolymers
with any given primary sequence. 
The kinetic equations earlier derived by us are transformed to
a form containing only the mean squared distances between pairs of
monomers. These equations are further expressed in terms of 
instantaneous gradients of the variational free energy.
The method allowed us to study exhaustively the
stability and conformational structure of some
periodic and random aperiodic sequences.
A typical phase diagram of a fairly long amphiphilic heteropolymer chain
is found to contain phases of the extended coil, the homogeneous globule,
the micro--phase separated globule, and a large number of frustrated
states, which result in conformational phases of
the random coil and the frozen globule.
We have also found that for a certain class of sequences the frustrated phases
are suppressed.
The kinetics of folding from the extended coil
to the globule proceeds through
non--equilibrium states possessing locally compacted, but partially
misfolded and frustrated, structure. 
This results in a rather complicated multistep
kinetic process typical of glassy systems.
\end{abstract}


\section{Introduction}\label{sec:intro}

Study of the conformational transitions of heteropolymers in dilute
solutions is important for many applications from
the chemical industry to biotechnology. 
Directed more towards the former,
there has been a significant amount of theoretical work
carried out on concentrated copolymer solutions, mixtures and blends 
\cite{Blends,deGennes,Cloizeaux,Edwards,GrosbKhokh,%
BlockCopolymers,Fredrickson,OrlandEtc,PanEruSha} 
using various types of the density formalism.
However, these approaches are not valid at infinitely low dilution where the
fundamental interactions of the individual macromolecule determine
its conformational state.  
This situation is more relevant for biochemistry.
The problem is even harder to address at
non--equilibrium conditions typical for biopolymers {\it in vivo}
\cite{Protein-book}.

For these reasons we have been working for some time on developing
an adequate statistical mechanical technique for studying the equilibrium
structure and kinetics across phase transitions in heteropolymers
\cite{GscBlock,GscRandom,GscRanPhd}. 
The main idea was to extend the Gaussian self--consistent (GSC) 
method, originally proposed for the homopolymer 
(see e.g. \cite{GscHomKin} and references therein), 
to the case of inhomogeneous 
monomer interactions. This has been achieved in Ref. \cite{GscBlock}
where we have derived the complete set of non--linear kinetic equations
for complex valued equal--time correlation functions of the Fourier transforms
of the monomer coordinates. There we have analysed in some detail
the simplest periodic (ab) copolymer, but study of more complicated
sequences remained out of our practical
computational reach. The relation of the
kinetic equations to the equilibrium free energy, as well as the expression
for the entropy, were also unknown to us at that stage. Thus, the phase
diagram have not been elucidated and certain other important
issues have not been addressed in that work.

In the following two works \cite{GscRandom,GscRanPhd} we have 
achieved further progress and resolved most of these questions, though
at the cost of loosing detailed information about individual
sequences. Namely, we have performed averaging of the GSC equations over
the quenched disorder in the monomer amphiphilic strength. 
This yielded a closed set of kinetic equations for description of
random copolymers with a Gaussian distribution of the disorder.
Such an idealised system is very interesting in itself, 
particularly since there
is a hope that its study could shed light on some features of the 
extremely complex problem of protein folding  \cite{Protein-book}.

To render the quenched disorder problem more tractable certain
perturbative approximation was necessary \cite{GscRandom}.
It became clear 
in Ref. \cite{GscRanPhd} that this
causes some deficiencies in the equilibrium limit of the
formalism, and we have alluded as to how these could be alleviated
in higher orders of the expansion.

Therefore, it seems important to revisit the 
general formalism of Ref. \cite{GscBlock} in order
to resolve remaining difficulties,
maintaining the rich information about particular sequences and avoiding any
further approximations. 
Furthermore, there exists, till now, no simple theoretical procedure
capable of giving the equilibrium conformational states of a
heteropolymer with arbitrary sequence, that can in a consistent manner
also give the full kinetic pathway between these states.
The work presented here achieves this.
To demonstrate the strength of the extended GSC method
we consider a number of interesting examples of heteropolymer
sequences.
Different kinds of interactions, such the hydrodynamic interaction,
may be straightforwardly incorporated into the scheme.
A rich and nontrivial physical picture emerges as a result of
this theoretical progress.

\section{The Bead--and--Spring Model} \label{sec:model}

Traditionally,
we proceed from the coarse--grained description of the polymer
chain \cite{deGennes,Cloizeaux,Edwards}
with the spatial coordinates ${\bf X}_n$ ascribed to the $n$-th monomer.
It is assumed that the long timescale evolution of conformational changes
is well represented by the phenomenological Langevin equation,
which upon neglecting the backflow effect of the solvent may be 
written as,
\begin{equation} \label{eq:Lang}
\zeta_b \frac{d}{dt} {\bf X}_n = - \frac{\partial H}{\partial {\bf X}_n}
+\bbox{\eta}_n(t),
\end{equation}
where $\zeta_b$ is the ``bare'' friction constant per monomer.
For the discussion of the hydrodynamic interaction we refer the reader
to Appendix \ref{app:hydro}.
The thermal fluctuations are incorporated via the Gaussian noise 
which, according to the Einstein law, is characterised by the second momentum,
\begin{equation}
\langle \eta^{\alpha}_n(t)\,\eta^{\alpha'}_{n'}(t') \rangle
=2k_B T \zeta_b\, \delta^{\alpha,\alpha'}\delta_{n,n'}\delta(t-t'),
\end{equation}
where the Greek indices denote the spatial components of 3-d vectors. 

In the current treatment the solvent is effectively excluded
from the consideration \cite{Footnote1} 
and the resulting monomer interactions
are described by the effective free energy functional,
\begin{equation} \label{eq:H}
H= \frac{k}{2}\sum_n ({\bf X}_n-{\bf X}_{n-1})^2+
\frac{\Kappa}{2}\sum_n({\bf X}_{n+1}+{\bf X}_{n-1}-2{\bf X}_n)^2 
+\sum_{J=2}^{\infty}\sum_{\{n\}}u^{(J)}_{\{n\}} \prod_{i=1}^{J-1}
\delta({\bf X}_{n_{i+1}}-{\bf X}_{n_1}),
\end{equation}
where in principle $u^{(J)}_{\{n\}}$ are allowed to have any
dependence on the site indices $\{n\}\equiv \{n_1, \ldots, n_J\}$.

The first two terms in Eq. (\ref{eq:H}) describe the connectivity
and the stiffness of the chain. Their coefficients have the following 
simple meaning:
$k=k_B T/l^2$ and $\Kappa=k_B T \lambda /l^3$ with $l$ and $\lambda$ called
the statistical segment length and the persistent length \cite{Edwards,Porod} 
respectively. Apart from these interactions, local along the chain, 
there are also long--range volume interactions represented by the
virial--type expansion \cite{deGennes,Cloizeaux}
in Eq. (\ref{eq:H}). The latter reflects
the hard core repulsion and weak attraction between monomers, but also the
effective interaction mediated by the solvent--monomer couplings $I_{n}$.

The coefficients in this expansion may be calculated as functions
of the temperature and the parameters of molecular interaction.
However, for our purposes we do not need to know their explicit form here
as we shall keep only a few first terms. Appropriate coefficients then 
may be viewed as independent phenomenological parameters which could
be directly related to experimentally measurable quantities. 

In Refs. \cite{GscBlock,GscRanPhd} we have discussed that the case of
amphiphilic heteropolymers, for which monomers differ only in the
monomer--solvent coupling constants, corresponds to the following choice of 
site--dependent second virial coefficients in Eq. (\ref{eq:H}),
\begin{equation}
u^{(2)}_{nn'}=\bar{u}^{(2)}+\frac{1}{2}(I_n+I_{n'}), \qquad \sum_n I_n=0. 
\end{equation} 
The mean second virial coefficient $\bar{u}^{(2)}$ is associated
with the quality of the solvent: positive values correspond to the good
solvent (where effective two--body 
repulsion of monomers results in the extended
coil conformation), and the negative values correspond to the bad solvent
condition (where the effective two--body attraction tends 
to compact the chain).

The set of couplings $\{I_n\}$ expresses the chemical composition,
or using the biological terminology, the primary sequence of a heteropolymer.
For simplicity, in the consequent sections we shall consider 
examples for which these constants can be parametrised in a simple way:
$I_n=\Delta\,\sigma_n$, where variables $\sigma_n$ can only take three values:
$-1,1$ or $0$ corresponding to (a) hydrophobic , (b) hydrophilic and 
(c) ``neutral'' monomers respectively. 
The parameter $\Delta$ is called the degree of amphiphilicity of the chain.   
Note that for more complicated than
binary sequences there is another relevant dispersion,
$(\Delta_\sigma)^2 = \frac{1}{N}\sum_m \sigma_m^2$, and the combination
$\Delta\Delta_\sigma$ is a more appropriate variable.

\section{The GSC kinetic equations}\label{sec:kineq}

In work \cite{GscBlock} we have derived a set (18) of the GSC
kinetic equations for the equal--time correlation functions of
the Fourier transforms of monomer coordinates ${\cal F}^{AA'}_q(t)$.
There, at the end of Sec. II, we have mentioned that a polymer with
no periodic structure may be described by choosing the number of blocks
$M=1$.  The equations for the correlation functions,
\begin{equation}
{\cal F}_{mm'}(t) \equiv \frac{1}{3}\biggl\langle {\bf X}_m(t)
\,{\bf X}_{m'}(t) \biggr\rangle,
\end{equation}
contain some redundancies and also
are intermixed with the diffusive degree of freedom describing the motion of
the centre--of--mass. Obviously,
for a single chain the latter can be easily decoupled
from the intra--molecular degrees of freedom by introducing the
mean squared distances between pairs of monomers,
\begin{equation} \label{eq:Ddef}
D_{mm'}(t) \equiv \frac{1}{3}\biggl\langle ({\bf X}_{m}(t)-{\bf X}_{m'}(t))^2
\biggr\rangle = {\cal F}_{mm}+{\cal F}_{m'm'} -2 {\cal F}_{mm'}. 
\end{equation} 
It turns out that from Eq. (18) in Ref. \cite{GscBlock} 
after certain algebraic manipulations one can obtain
a closed set of equations for the quantities $D_{mm'}(t)$.

Taking into account the additional bending energy contribution in Eq. 
(\ref{eq:H}), the GSC equations may be written down
in the most general form as follows,  
\begin{eqnarray} \label{eq:Dkineq}
\frac{\zeta_b}{2}\frac{d}{dt} D_{mm'} &=& 2k_B T(1-\delta_{mm'})
+k(D_{mm',m+1\,m}+D_{mm',m-1\,m}+(m\leftrightarrow m')) \nonumber\\
&&-\Kappa (D_{mm',m+2\,m+1}+D_{mm',m-2\,m-1}
- 3D_{mm',m-1\,m}-3D_{mm',m+1\,m}+(m\leftrightarrow m')) \nonumber \\
&&+\sum_{J=2}^{\infty}\sum_{\{n\}}
\frac{\hat{u}^{(J)}_{\{n\}}}{({\rm det}\Delta^{(J-1)})^{5/2}}
\sum_{i,j=1}^{J-1}\bar{\Delta}_{ij}^{(J-1)}
(\delta^{m}_{n_1}+\delta^{m'}_{n_{i+1}}
-\delta^{m}_{n_{i+1}}-\delta^{m'}_{n_1}) D_{mm',n_1n_{j+1}},
\end{eqnarray}
where we have introduced the four--point correlation functions, $D_{mm',nn'}$,
and the matrix $\Delta^{(J-1)}_{ij}$ of size $(J-1)$ with the cofactor
$\bar{\Delta}^{(J-1)}_{ij}$,
\begin{eqnarray}
D_{mm',nn'} &\equiv& -\frac{1}{2}(D_{mn}+D_{m'n'}
-D_{mn'}-D_{m'n}), \label{D4} \\
\Delta^{(J-1)}_{ij} &\equiv& D_{n_1 n_{i+1},n_1 n_{j+1}}.
\end{eqnarray}
Similarly to Eq. (31) in Ref. \cite{GscBlock} for the mean energy
${\cal E}= \langle H \rangle$ we have,
\begin{eqnarray}
{\cal E} &=& \frac{3k}{2}\sum_n D_{n\,n-1,n\,n-1}
+\frac{3\Kappa}{2}\sum_n(2 D_{n+1\,n,n+1\,n}+2 D_{n-1\,n,n-1\,n}
-D_{n-1\,n+1,n-1\,n+1}) \nonumber\\
&&+\sum_{J=2}^{\infty}\sum_{\{n\}}\hat{u}^{(J)}_{\{n\}}({\rm det}
\Delta^{(J-1)})^{-3/2}. \label{eq:E}
\end{eqnarray}

It is interesting to note that in the case of the homopolymer 
the  right--hand side of the
kinetic equations (\ref{eq:Dkineq}) may be rewritten via
the instantaneous gradients of the variational free energy
by introducing the normal modes
(see Eq. (5) in Ref. \cite{GscTor}). This establishes connection
of the stationary limit of our kinetic approach with the equilibrium
theory, in which we recover the Gibbs--Bogoliubov variational principle.
We also note that if first order phase transitions are involved, 
one has to possess the expression for the free
energy in order to determine the phase boundaries by finding the
global free energy minimum. 

The variational free energy, ${\cal A}$, based on the Gaussian {\it Ansatz}
for equal--time pair correlation functions, contains two terms, 
${\cal A} = {\cal E} - T{\cal S}$. Naturally, the 
mean energy term is given by Eq. (\ref{eq:E}). 
The second, entropic contribution, is calculated in Appendix \ref{app:var}.
Let us summarise the result of that calculation here. 
One representation for the entropy that is suitable for numerical analysis is,
\begin{equation} \label{Sr}
{\cal S}=\frac{3}{2}k_B \log {\rm det}\, R^{(N-1)},
\end{equation}
where we have used the determinant of the $(N-1)$-dimensional major
submatrix of the matrix,
\begin{equation} \label{RtoDD}
R_{nn'}=\frac{1}{N^2}\sum_{mm'}D_{nm,n'm'}
=-\frac{1}{2}\Lambda^0 \bullet D \bullet \Lambda^0,
\end{equation}
and the matrix $D$ obviously has the elements equal to $D_{nn'}$.
The reason for appearance of the truncated matrix is that we have excluded
the zero eigenvalue of $R$ related to the translational
invariance.
Here we have also introduced the $(N-1)$-dimensional 
orthogonal projector $\Lambda^0$ such that,
\begin{equation}
(\Lambda^0)^2=\Lambda^0, \quad
(\Lambda^0)^T=\Lambda_0, \quad
\sum_{n} (\Lambda^0)_{nk}=0,
\end{equation}
with the matrix elements $(\Lambda^0)_{nk}=\delta_{n,k}-1/N$.
This matrix has obviously one zero eigenvalue and $N-1$ degenerate
eigenvalues equal to $1$.

However, for analytical treatment it is more convenient to obtain a slightly
different expression by regularising the zero eigenvalue (i.e. by imposing
the constraint $\sum_n \bbox{X}_n=0$ as a ``soft'' condition in Eq.
(\ref{eq:Tr})),
\begin{equation} \label{Se}
{\cal S} =\frac{3}{2}k_B \lim_{\epsilon\to 0} 
\log  \frac{{\rm det} (R+\epsilon \bbox{1}) }{\epsilon}.
\end{equation}
Since the matrix $(R+\epsilon\bbox{1})$ 
is nondegenerate we can easily differentiate Eq. (\ref{Se}) so that,
\begin{equation} \label{eq:Sder1}
Q_{ik} \equiv \lim_{\epsilon\to 0}
\sum_j D_{ij} \frac{\partial}{\partial D_{kj}} {\rm Tr}\,
\log (R+\epsilon\bbox{1})=\sum_j D_{ij} {\rm Tr}_0 \left( 
\frac{\partial R}{\partial D_{kj}} \bullet R_0^{-1}
\right)=
(\Lambda^0)_{ik},
\end{equation}
where inside the trace ${\rm Tr}_0$ over the $(N-1)$-dimensional subspace
projected out by $\Lambda^0$ the matrix $R$ becomes invertible with the inverse
denoted as $R_0^{-1}$.
This allows us to obtain the combination which appears in the kinetic equations,
\begin{equation} \label{eq:Sder}
Q_{ii}+Q_{kk}-Q_{ik}-Q_{ki}=2(1-\delta_{i,k}).
\end{equation}

These preliminaries are sufficient to prove the desired relation.
Indeed, using Eq. (\ref{eq:Sder}) by direct and tedious differentiation of
Eq. (\ref{eq:E}) one can show that the kinetic equations 
(\ref{eq:Dkineq}) may be expressed through the instantaneous
gradients \cite{Footnote2}
of the variational free energy as,
\begin{equation} \label{eq:kinmain}
\frac{\zeta_b}{2}\frac{d}{dt}D_{mm'}(t) = -\frac{2}{3}
\sum_{m''}(D_{mm''}(t)-D_{m'm''}(t))\left(
\frac{\partial {\cal A}[D(t)]}{\partial D_{mm''}(t)}-
\frac{\partial {\cal A}[D(t)]}{\partial D_{m'm''}(t)}
\right).
\end{equation}
This formula, together with Eqs.
(\ref{eq:E},\ref{Se}), is the key formal result of the current work.
The structure of Eq. (\ref{eq:kinmain}) is sufficiently
nontrivial to be guessed from phenomenological arguments
and has been derived in a systematic manner proceeding from 
Eq. (\ref{eq:Dkineq}). 

We would like to comment here that although, for simplicity, we have presented
the explicit formulae above for a ring polymer, our current formalism 
is general and covariant. In fact, the kinetic equations (\ref{eq:kinmain})
are valid for any topology of the chain. Thus, it is
straightforward to consider more complicated topologies
such as a star, brush, network, branched chain and so on. For that
it is sufficient to modify only the spring and stiffness terms
in Eq. (\ref{eq:H}), and respectively in Eq. (\ref{eq:E}).
For example, to describe an open chain one has simply to suppress
the term with $n=0$ in the connectivity contribution,
and the terms with $n=0,N-1$ in the stiffness contribution to the energy.
We undertake a detailed comparison of ring versus open homopolymers
in kinetics at the collapse transition of the homopolymer
in a separate work \cite{GscOpen}.
Another interesting possibility is that the general equation (\ref{eq:kinmain})
is also valid for models with different ways of representing the connectivity
and stiffness. For example, the freely rotating model 
\cite{Edwards}  can be obtained by suppressing the first two terms in
Eq. (\ref{eq:E}) and instead keeping fixed the following mean squared
distances: $D_{m,m+1}=b_0^2$ and 
$D_{m,m+2}=4b_0^2 \sin^2 \theta/2$, where $b_0$ and $\theta$ are
the bond length and angle. This is easy to prove by adding
appropriate ``soft'' constraints to the free energy functional
and taking the consequent limit in Eq. (\ref{eq:kinmain}),
so that they become delta--functions in the partition function.

Finally, let us introduce two main observables: the mean squared radius 
of gyration and the degree of micro--phase separation \cite{GscRandom},
\begin{equation}
R_g^2=\frac{1}{2N^2}\sum_{mm'} D_{mm'}, \quad
\Psi = \frac{1}{N^2\,R_g^2\, \Delta\Delta_\sigma}
\sum_{mm'}(u^{(2)}_{mm'}-\bar{u}^{(2)})
D_{mm'}.
\end{equation}
The second parameter has a meaning of the dimensionless correlation of
the matrices of the relative two--body interaction and the
mean squared distances. 
For heteropolymers with two types of monomers it
characterises the difference between the mean squared radii of
gyration of hydrophilic and hydrophobic species, and for the symmetric
composition with their numbers equal we have a simple relation: 
$\Psi=(R_g^2(b)-R_g^2(a))/(2R_g^2)$.

\section{Numerical results} \label{sec:num}

The self--consistent kinetic equations (\ref{eq:kinmain}) have been
studied numerically using the explicit formulae 
(\ref{eq:c1},\ref{eq:c2},\ref{eq:c4}) for the effective
potentials (see Appendix \ref{app:der}), and the expression given by the first 
term in the right--hand side of Eq. (\ref{eq:Dkineq}) 
for the entropic contribution.
We used the fifth order adaptive step Runge--Kutta method
\cite{NumerRecip} to improve stability of the solution which, for
large amphiphilicity parameter, $\Delta$, is characterised
by a rather rugged free energy landscape.
We note that such kinetic method for finding the equilibrium 
distributions is more reliable and efficient than the standard
methods of free energy minimisation if there are lots of mountains
and valleys on its surface.  
We also refrain from study of the influence of the hydrodynamics in this
paper for the analysis and results are complicated enough already.
Besides, the hydrodynamics in the preaveraged approximation does
not affect the equilibrium state, which is recovered by taking the
stationary limit in the kinetic equations.

We include the volume interactions up to the three--body terms only,
i.e. $u^{(J)}_{\{n\}}=0$ for $J > 3$.  As can be seen from
Eqs. (\ref{eq:kinmain},\ref{eq:c2})
the computational time per time step scales
with the chain length as $t_c \sim N^3$ here. This performance is intermediate
between that of the homopolymer $t_c \sim N^2$ \cite{GscHomKin,GscTor} and
that of the random copolymer $t_c \sim N^4$ \cite{GscRandom,GscRanPhd}.
The performance of the formalism in Ref. \cite{GscBlock} was
$t_c \sim K^4 M^2$, where $K$ and $M$ are the block length and number of blocks
respectively. Besides, that formalism relied on the use of complex
variables, and the unitary transformation to the real basis was not
an easily automatised task for complicated sequences. Moreover, the treatment
of the diffusive mode was nontrivial and sequence dependent.
Thus, in every respect, the current scheme is most attractive 
for study of heteropolymer sequences from the computational point of view.

It is natural to work here with combinations ${\cal L}=(k_B T/k)^{1/2}$ and
${\cal T}=\zeta_b/k$ as the units of size and time. We choose $k=1$,
$k_BT =1$ and $\zeta_b=1$ to fix ${\cal L}$ and ${\cal T}$ equal
to unity.
In addition, we fix the following interaction parameters:
the third virial coefficients $u^{(3)}_{mm'm''}=10\ k_B T {\cal L}^6$ and the
stiffness $\Kappa=0$ as we did in Ref. \cite{GscBlock}.

Now, we turn to discussion of concrete results.
We have studied ring chains of the lengths $N=30,\ 60,\ 90$.
We shall present here the most complete results for $N=60$ and discuss
the $N$ dependence only briefly. We have
examined many different
sequences. However, we present our data in this paper only 
for three particular choices which
we found most typical and illustrative of the heteropolymer behaviour.
These are chains of: 30 (ab) blocks,  
10 (aaabbb) blocks and 2 (abacbbcabccbcaaacbcccbbaacbcca) blocks,
which we call the ``short'' blocks, ``long'' blocks and ``random''
sequences respectively (see the end of Section \ref{sec:model} for
the monomer notations).

\subsection{Equilibrium phase diagrams} \label{subsec:equ}

The phase diagrams in terms of the mean second virial coefficient
$\bar{u}_2$ and the amphiphilicity $\Delta$ for the above mentioned 
three sequences are presented in Figs.~\ref{fig:1}-\ref{fig:3}.
For positive values of $\bar{u}^{(2)}$ and comparatively 
small values of $\Delta$
the conformational state of the chain is akin to a homopolymer  
extended coil (see Fig.~\ref{fig:dmneq}a). 
By decreasing $\bar{u}^{(2)}$ to the negative region the
chain is caused to undergo a continuous (second--order--like) collapse
transition (curve (C)), that is characterised by a rapid fall of the radius of
gyration \cite{Footnote2b} (see Fig.~1 of
Ref. \cite{GscHomKin}) and the change of the fractal dimension. 
Proceeding from the collapsed globule 
at a fixed negative $\bar{u}^{(2)}$ the
increase of the amphiphilicity $\Delta$ also causes a continuous transition 
(curve (S))
to the micro--phase separated (MPS) globule
\cite{Footnote3}. During this transition the
system size, $R_g$, monotonically increases (see Fig.~\ref{fig:4}) 
and the mean energy ${\cal E}$ decreases
in agreement with our earlier results in Refs. \cite{CoplmMonte,GscBlock}.
The former change is more pronounced for the ``long'' blocks compared to other
sequences, for which the connectivity constraints impede formation of
structures with a hydrophilic shell and hydrophobic core.
The MPS order parameter $\Psi$ (see Fig.~\ref{fig:5})
increases almost linearly for small $\Delta$,
then after the transition asymptotically saturates.

The transition from the coil at large values of $\Delta$ to the 
MPS globule turns out to be more complicated, and essentially 
dependent on the sequence. In case of the ``long'' blocks (Fig.~\ref{fig:2})
the collapse
transition to the MPS globule becomes discontinuous (first--order--like).
Thus, inside the boundaries of metastability, designated by spinodals 
I' and I'', there are two competing minima of the free energy: the coil
and the MPS globule. The former minimum is characterised by a large
size, $R_g$, (on the right of Fig.~\ref{fig:6}) and a small MPS order 
parameter $\Psi$ (Fig.~\ref{fig:7}), while for the latter minimum the situation is
reversed (see the left hand side of Figs.~\ref{fig:6},\ref{fig:7}).
The depths of the free energy minima become exactly 
equal on the transition curve I in Fig.~\ref{fig:2}.

For the ``short'' blocks (Fig.~\ref{fig:1}), as well as for many 
random sequences 
(such as in Fig.~\ref{fig:3}), the phase diagram is much more complicated.
Starting from some value of $\Delta$ and for
intermediate values of $\bar{u}^{(2)}$ there appear additional solutions
corresponding to local minima of the free energy. The broad region
where this could take place is bounded by the curves I' and II'' 
in Figs.~\ref{fig:1},\ref{fig:3}.
With increasing $\Delta$ the number of such solutions grows quickly.
Significantly, in a region of the phase diagram some of these become the
main free energy minimum. As the number of such solutions grows 
roughly exponentially with the chain length, we do not attempt to draw
all their boundaries of (meta)stability. Instead, we shall designate
them collectively as {\it frustrated} phases, explaining this terminology
below.

An important point here is that, as our analysis shows, 
these solutions become dominant 
in a narrow region of the phase diagram
due to a subtle competition between the mean energy and the entropy.
The MPS globule is entropically unfavourable there because the
overall shrinking force is insufficiently strong.
The values of $R_g^2$ and $\Psi$ are
intermediate for these solutions and lie between those of the coil and
MPS globule (see Figs.~\ref{fig:8},\ref{fig:9}). 
In this sense, we can call them nonfully
compacted and misfolded states. In comparison, the MPS globule 
(see Fig.~\ref{fig:dmneq}d) has a
more compact size and better optimised volume interactions, what is manifested
in a higher value of $\Psi$.     

Most interesting is the local structure of these additional phases.
Let us discuss the particular example of the 30(ab) sequence.
The formalism of Ref. \cite{GscBlock} has used heavily the assumption of
certain symmetries for the mean squared distances $D_{mm'}$ due to which
these variables
can take only $3M$ independent values, where $M$ is the number of blocks.
These symmetries are: the block translation invariance,
\begin{equation} \label{eq:sym}
D_{m+Ki,m'+Ki}=D_{m,m'}, \quad  \mbox{for any
$m,m'$ and $i$}, 
\end{equation}
where $K$ is the block length, and the more
complicated inversion symmetry discussed in detail in Ref. \cite{GscBlock}.
These symmetries have a simple meaning --- a renumbering
of monomers does change the average properties over the
statistical ensemble
--- the interactions of a ring chain remain the same.
However, the maximal possible number of dynamical variables in the GSC method
is much larger and equals to $N(N-1)/2$.
Surprisingly,
it turns out that the frustrated phases are characterised by
spontaneous breaking of these symmetries, and therefore only
the current version of the method that takes into account
all degrees of freedom can describe them.
Thus, the property (\ref{eq:sym}) is no longer valid.
This phenomenon
describes formation of local frustrated heterogeneities 
(see Figs.~\ref{fig:dmneq}b,c)
in which pieces of the chain form MPS clusters \cite{CoplmMonte}
that are prevented from further
coalescing by their hydrophilic shells and high entropic barriers.

The role of spontaneous symmetry breaking is well recognised in 
equilibrium statistical mechanics. 
What is striking here is that the
number of distinct spontaneously broken states becomes huge for
large system sizes. This diversity and a 
special foliating structure of
various branches leads in the thermodynamic limit to what is known as
a spin glass like frozen phase 
\cite{Mezard-book} of random copolymers 
(see e.g. Refs. \cite{GscRandom,GscRanPhd} and numerous references therein). 
We shall return to the issue of spontaneous symmetry breaking 
in kinetics in subsection \ref{subsec:kin}.

Even though the kinematic symmetries are not present from the outset
for arbitrary, or random, sequences,
the structure of the phase diagram (Fig.~\ref{fig:3})
and behaviour of main observables (Figs.~\ref{fig:8},\ref{fig:9})
remain very similar.
It is the particular structure, number and 
the shape of boundaries of frustrated phases
that are very sensitive to the sequence.
The symmetry that may be broken in this case has a subtler, dynamic, meaning
and may be expressed in terms of the replica formalism \cite{Mezard-book}.
In a sense, for a very long periodic chain, blocks may be viewed as
identical copies of a smaller block--length chain. Thus, it is by no means
surprising that the replica symmetry breaking in our case  for periodic systems
takes such an explicit
manifestation in the  breaking of the block translational symmetry.
This important point was completely missed, nor could it be discovered,
in our considerations of Ref. \cite{GscBlock}.
Therefore we have achieved here a new significant insight into the problem 
by a simple extension of the GSC method.

An interesting feature of our phase diagrams is that the region between
spinodals I' and II'', designating where the frustrated phases can exist,
expands dramatically with increasing chain length.  
For example, at $\Delta=25$, for the 15(ab) sequence  it lies approximately 
between $-19 <\bar{u}^{(2)} <10$, while for the 30(ab) ---
between $-39 < \bar{u}^{(2)} < 11$,
and the curve II'' goes nearly vertically downwards for larger
$N$. 
According to the above interpretation, for infinitely long chain 
somewhere in this broad region and
close to its boundaries there are the actual glass transition curves,
which should be determined using proper glass order parameters.
Thus, for short chains the curves I' and II'' may be viewed as
approximate indicators of the freezing transitions.
The former distinguishes between the homopolymer--like and the 
random coil, while the latter --- between the homopolymer--like
(liquid) and the frozen globules.
As for the region of stable MPS globule, it gets relatively smaller for
larger systems. That is not surprising ---
the frustrated phases expand and the phase separation involving larger
spatial scales requires stronger interactions.
Having understood the identifications for the spinodals, 
we now can recognise  in Figs.~\ref{fig:1},\ref{fig:3} the main features of
the phase diagram, though rather distorted, 
of random copolymer model presented in Ref. \cite{GscRanPhd}.

Finally, let us comment on the phase diagram of the ``long'' blocks 
(Fig.~\ref{fig:2}).
The micro--phase separation is obviously easier in this case and 
it dominates for large values of $\Delta$, so that 
the frustrated phases are suppressed.
We found that, qualitatively, in order to form a frustrated phase,
in a finite range of $\Delta$ values, 
the number of (not necessarily identical)
pieces of the chain with competing interactions should be larger
than some critical number, in principle, weakly dependent on $N$.
Here are a few examples conforming to this qualitative criterion: 
the phase diagram of 10(ab) behaves roughly as for 10(aaabbb),
but for 15(ab) it behaves similarly to 30(ab); 
while for 15(aabb) the phase
diagram looks like for 15(aaabbb) at small and moderate values of 
$\Delta$, it becomes as for 30(ab) for much larger values of the 
amphiphilicity.
It is reasonable to conjecture therefore that for a large number of (aaabbb)
blocks, as well as for extremely high values of 
$\Delta$ and just 10 blocks, the frustrated phases may be found again.

\subsection{Folding kinetics} \label{subsec:kin}

Here we shall consider the time evolution of the conformational
state of the system away from its initial equilibrium 
after it has been subjected to an instantaneous
temperature jump that causes the two--body interaction parameters
$\bar{u}^{(2)}$ and $\Delta$ to change. We are interested in quenches
from the homopolymer coil, where all monomers are equally 
hydrophilic ($\bar{u}^{(2)} >0$ and $\Delta=0$), to the region of parameters
corresponding to the MPS globular state, so that the `a' species became
strongly hydrophobic and the `b' species remained weakly hydrophilic
($\bar{u}^{(2)} \ll 0$ and $\Delta \gtrsim |\bar{u}^{(2)}|$).

The temporal behaviour of the mean squared radius of gyration,  $R_g^2(t)$, 
the MPS parameter, $\Psi(t)$, and the instantaneous free energy,
${\cal A}(t)$, in kinetics of folding for different sequences is
presented in Figs.~\ref{fig:10}-\ref{fig:12}. 
For the homopolymer (curves (A)) $R_g^2$ and ${\cal A}$ decrease
monotonically to their final equilibrium values, while the MPS
parameter, $\Psi$, remains identically zero for there is no distinction between
different monomer species. These curves agree with the earlier
results of Ref. \cite{GscHomKin} and serve for reference purpose here.
 
Now let us discuss the curves (B) corresponding to the periodic (ab) sequence.
In the previous subsection we have mentioned that the current formalism 
yields results consistent with those of Ref. \cite{GscBlock} only
beyond the parameter region of the frustrated phases, which are 
characterised by  spontaneous breaking of the block symmetry.
In kinetics the situation is somewhat similar, but the consistence with
the previous simplified formalism is even more limited.
Really, the mean squared radius of gyration remains close to that of the 
``effective'' homopolymer (curve (A)) during the first
kinetic stage (see Eq. (74) in Ref. \cite{GscBlock}).
Interestingly, this is not quite so for more complicated sequences.
The MPS parameter, $\Psi(t)$, for the (ab) copolymer grows slowly in a way 
similar to the splitting of the Fourier modes 
${\cal F}^{11}_q(t)-{\cal F}^{00}_q(t)$
for large indices $q$ (see Fig.~4 in Ref. \cite{GscBlock}).
However, contrary to the strange conclusion 
in Ref. \cite{GscBlock} that
the kinetics of the  (ab) copolymer proceeds faster than for the homopolymer, 
we now have a, natural from the point of view
of Monte Carlo simulations \cite{CoplmMonte}, 
slowing down of copolymer kinetics.
Analysis of $D_{mm'}(t)$ shows that
this effect is entirely due to the spontaneous breaking of the block
symmetry in kinetics, something that has not been accounted for in 
Ref. \cite{GscBlock}.
Indeed, in Fig.~\ref{fig:13} we 
exhibit the time dependence of the mean squared
distances between two nearest hydrophobic monomers $D_{2k,2k+2}(t)$
for the (ab) copolymer in kinetics after a quench to the MPS globule region.
For early, as well as for late, times these functions for different $k$ 
are exactly equal to each other. However, there is well defined intermediate
period in time where the block symmetry is spectacularly broken
(see Fig.~\ref{fig:dmnkin}).
We remark that the symmetry breaks and restores in a step--like manner
(e.g. $D_{0,2}$ and $D_{4,6}$ join together at $t\simeq 10$, 
earlier than with other functions) and also that this effect is 
relatively strong. 

We would like to make a general comment here on the nature of
spontaneous symmetry breaking in kinetics.
Thus, normally in such situations at equilibrium there exist
a thermodynamically unstable symmetric free energy minimum and a
disjoint set of symmetry broken minima, which may be transformed
to each other by the residual subgroup of symmetry transformations.
These states may also be obtained kinetically as infinite time limits 
of the time evolution starting from any, for example, the symmetric 
initial state, which happens to be the main free energy 
minimum before the quench.

However, the formal structure of the GSC 
kinetic equations (\ref{eq:kinmain}) is
such that they yield a symmetric solution at any moment in time provided
one proceeds from the symmetric initial condition. 
A question arises then --- 
how can one obtain the kinetics that could lead eventually to the
multitude of final states with broken symmetry?
The answer is clear in the exact theory --- 
there are fluctuations that can transform between
different spontaneously broken states in kinetics.

The GSC method presents, though an optimised and improved, but
still a mean field type theory,
where such fluctuations are not properly included. 
Manifestation of the
kinetic spontaneous symmetry breaking takes a different form there.
Namely, at some moment in time the symmetric solution of the kinetic equations 
becomes unstable with respect to perturbations (whether of the initial 
condition, or of the interaction 
matrix, $u^{(2)}_{nn'}$). Thus, for example, one can add an infinitesimal
symmetry breaking term $\varepsilon_{nn'}$
to the two--body interaction matrix and consider the limit
of vanishing perturbation in the solution. 
Different choices of the form of $\varepsilon_{nn'}$ 
in the unperturbed limit
would yield  all possible types of spontaneously broken
kinetic evolution, which are, of course, equivalent to each other.
Numerically, such a regularisation procedure is not even necessary.
There is always an intrinsic perturbation 
due to computer round off and numerical integration errors. 
Thus, if the symmetry is favourable to be kinetically broken somewhere, 
numerically one obtains one of the spontaneously broken solutions there,
rather than the unstable symmetric solution, unless the symmetry
conditions have been imposed by hand. 
In this situation improvement of the 
numerical precision would have no profound effect --- 
the kinetic process either does not change, or
can change only up to an experimentally unobservable 
residual symmetry transformation.

We note also that
such a procedure of perturbing the interaction matrix here is
analogous to introducing an external magnetic field in the Ising model.
The spontaneous magnetisation of a macroscopic sample of ferromagnet 
may be achieved
by gradually switching off the external magnetic field.
In the absence of the field
there would remain domains with long--range order,
but varying directions of the spontaneous magnetisation
canceling each other. Unfortunately, in our case it is not evident
how to experimentally implement an analogous gradual switching off of
$\varepsilon_{nn'}$.

Now returning to our results,
in Fig.~\ref{fig:10} one can see that the 
folding kinetics for the (ab) copolymer is 
about 3 times slower than for the homopolymer of the same length,
the effect being even stronger for other sequences we have considered. From 
Fig.~\ref{fig:11} and the phase diagrams in Figs.~\ref{fig:1}-\ref{fig:3} 
it is evident that
the considered quench results in the final state of  the MPS globule 
for the ``long'' blocks (curve (C)), whereas in the frustrated phases for two
other sequences (curves (B) and (D)). For the latter phases the 
MPS parameter is smaller and the radius is larger than for the MPS globule,
as we already know from the equilibrium considerations.
It is interesting to note that
the final relaxation to the frustrated phases may be rather unusual.
For example, in Figs.~\ref{fig:10},\ref{fig:11} for the ``random'' sequence  
$R_g^2(t)$ increases and 
$\Psi(t)$ decreases during the last kinetic stage,
what is the converse to the behaviour at final relaxations in other cases.
Another unusual observation is that
for some sequences the parameter $\Psi$ may even become negative
during some time in kinetics, something we never observed at equilibrium.
This shows that the structure of non--equilibrium conformations
can be very complicated.

The instantaneous free energy ${\cal A}(t)$ depicted in Fig.~\ref{fig:12}
turns out to
be the quantity most sensitive to the conformational structure of the
non--equilibrium state. From that figure it strikes that, while
the homopolymer folding is a single smooth relaxational process, the folding
of heteropolymers proceeds through a multistep acceleration--deceleration
process. The flat regions of a staircase--like function correspond
to temporary kinetic arrest of the system in transient non--equilibrium 
(mostly symmetry broken) conformations. Generally speaking, 
in the GSC method we deal with the time evolution of a statistical
ensemble of various initial conformations. The flat regions appear due 
to transient trappings of various members of the ensemble in their local
shallow energy minima. Since such minima are encountered at
different moments in time for different members of the ensemble,
their influence on the overall time evolution of averaged observables
is manifested in a smooth characteristic slowing down.

Qualitatively then, summarising the data for various
sequences, most of which we have suppressed here, 
we can say that the number of local minima 
affects the number of stair steps,
whereas the depth of minima determines the lengths of the steps. 
This interpretation is highly supported by the strong dependence
of the stair case structure on the chain length, the sequence and  the
interaction parameters, for it is known how the frequency and depths
of local potential energy minima depend on these factors.
On the one hand, it is known that
for a long heteropolymer chain the number of local
minima grows exponentially with the chain length $N$, and
indeed we see that 
the number of steps in ${\cal A}(t)$  grows just as quickly.
On the other hand, increasing of the amphiphilicity, $\Delta$, leads 
to higher depths of local minima as well as an increase in their number,
and really the steps in ${\cal A}(t)$ become longer and more numerous.
Let us give another example of such connection.
For the ``long'' blocks the interactions are less frustrated than, say, 
for the (ab) blocks,
thus there are much fewer of local energy minima, but 
the barrier between the coil and MPS globule is a higher.
As a result, the kinetic process
(curve (C) in Fig.~\ref{fig:12}) proceeds through only one 
rather long kinetically arrested step.

It is worthwhile to make a comment here on the notion of kinetic 
stages introduced
in earlier works \cite{GscKinet,GscHomKin,CoplmMonte}.
Those we associated with typical structures of the conformation
and accompanying kinetic laws. We distinguished at least the
following kinetic stages at the collapse of the homopolymer: 
early time necklace formation, middle time coarsening and 
a number of final relaxational processes.
The multistep character of folding observed in this work 
affects the middle kinetic stage, resulting in its considerable
complication and splitting into multiple substages with respective
complex kinetic laws that are determined by the sequence.
Universality of such kinetic laws is doubtful, but 
probably it can be recovered by
averaging over certain classes of sequences with similar folding properties.

Apparently, in the GSC method the kinetics is a motion 
in the space of $N(N-1)/2$ averaged dynamic variables, $D_{nn'}(t)$,
and it is determined by the profile of the free energy.
It may be instructive,  using Figs.~\ref{fig:10} and \ref{fig:12}, to present
the kinetics via  
a parametric plot of ${\cal A}$ vs $R_g$, the latter being the main,
though not the only,
relevant ``coordinate'' of the system.
That would produce a  kind of ``bottleneck'' picture that was much discussed 
by P.~Wolynes and others \cite{Wolynes-B} 
in relation to the  protein folding problem. 
This indicates that our method produces behaviour
that permits interpretation in terms of phenomenological 
energy landscape models.    

Finally, let us utilise once again
the connection between the kinetic evolution of the free energy and the
ruggedness of the potential energy landscape.
This would allow us to shed some light on the 
general structure of the energy landscape for
complex heteropolymers. Thus, a typical example of the ``random'' sequence
kinetics (curve (D) in Fig.~\ref{fig:12}) shows first a fast drop, 
then appearance of short
steps that are growing longer with time, until the last step becomes infinite.
This translates to the following equivalent energy landscape
versus some collective coordinate:
first there is a rapid drop away from the unstable coil state, then the
surface  flattens and small wrinkles appear on it, they are
growing larger in the amplitude gradually becoming high mountains and 
deep valleys, until there is eventually a very deep ravine corresponding to 
the ``ground'' state of the system separated by a very high barrier from
other minima. In our view,
the latter picture bears a remarkable resemblance to
a typical mean field energy landscape in spin glass systems 
\cite{Mezard-book} recently 
discussed by G.~Parisi \cite{Parisi-talk}. 
This observation seems encouraging to us and indicates that
the GSC method is capable
of describing very complex systems, though it is too detailed 
and expensive in the complete form for description on a macroscopic scale.
Nevertheless, by taking the quenched disorder averaging (similarly to Ref. 
\cite{GscRandom}) it is possible to achieve an adequate
description for heteropolymers that is alternative to the replica
formalism. 
We believe that the underlying connection between both
types of approaches has been clarified to some extent by the current work.

\section{Conclusion} \label{sec:concl}

In the present work we have extended the Gaussian self--consistent method
to permit study of heteropolymers with complicated primary sequences.
This has been achieved by transforming the kinetic GSC equations,
earlier derived for the Fourier modes,
to the form containing only the mean squared distances between
pairs of monomers and relating these equations to 
the instantaneous gradients of
the variational free energy calculated based on the Gibbs--Bogoliubov 
principle. 
The revised GSC formalism possesses important computational
advantages, but also it is fundamentally superior to its predecessor in that 
it can describe phenomena of spontaneous symmetry breaking and
formation of structural heterogeneities.

We then have applied the extended GSC method to some 
particular amphiphilic heteropolymers.
The equilibrium phase diagrams for these have been obtained in
a systematic manner. Apart from the coil and two globular states
we have discovered that in a wide intermediate region of the phase diagram
there may be a large number of frustrated partially misfolded
states.  Despite the fact that such states, the number of which grows 
exponentially with the chain length, 
are mostly metastable, some of them become the
dominant state in their rather narrow domains. The corresponding potential
energy profile of the system has strong resemblance to that of a
typical spin glass system. Thus, we may conclude that the
transition to these states in the thermodynamic limit corresponds to a glassy
freezing transition. 

We note that an observation that for sufficiently long 
sequences of alternating monomers (``short'' blocks) the 
micro--phase separated state is displaced by the region of glassy
frustrated phases has been known for quite some time and well understood
in the framework of the density theories \cite{PanEruSha}
as a result of the density fluctuations which modify
the mean--field behaviour. The block translational symmetry breaking in
our approach is another manifestation of the glassy phenomena earlier
extensively studied for melts of random heteropolymers.
In addition, we can see how the destruction of the micro--phase
structures occurs for finite--sized systems and how exactly it depends on a 
particular sequence.

The folding kinetics is found to be strongly affected by the
presence of these transient frustrated states along the kinetic pathway.
This leads to a complicated kinetic process consisting of multiple
steps with pronounced slowing down and then acceleration in the folding rate.
It is interesting to note that a typical fairly heterogeneous
chain with weakly hydrophilic and strongly hydrophobic units
folds kinetically first
to one of the misfolded states and can undergo a consequent
nucleation process to the main thermodynamic state of the micro--phase
separated globule.

These results for equilibrium and kinetics confirm many predictions
of the earlier treatment adopted for random heteropolymers in
Ref. \cite{GscRandom} based on a quasi--perturbative averaging of
the GSC equations over the quenched disorder. The latter approach
describes a large set of random heteropolymer sequences 
with a Gaussian distribution
of the monomer amphiphilicity. The kinetics of folding seems to be 
consistent between the previous and current approaches with respect to
main observables such as the radius of gyration, energy and the MPS
parameter. 
A smooth slow glass--like relaxation in the former approach is produced
by averaging over all different sequences with their particular
multistep processes of passage through the frustrated states.
Although here we observe strong dependence of the phase boundaries in
the equilibrium phase diagram on the primary sequence, the major
conclusion of Refs. \cite{GscRandom,GscRanPhd} that there are three 
distinct globular states of the homopolymer--like (liquid), 
the frozen (glassy) and the micro--phase separated (folded) globules,
and an additional state of the random coil, as well their relative location
in the diagram,
remain valid. We have previously discussed that
the equilibrium limit of the theory in the treatment of Ref. \cite{GscRanPhd}
had certain deficiencies due to the additional perturbative approximation,
which we attempted to remedy phenomenologically.
The current fundamental scheme is free of any such problems 
and provides a reliable test ground for development of further approximations.

We have also discovered that heteropolymer sequences of equal length 
can be roughly divided into distinct 
classes possessing similar phase diagrams and kinetic folding properties.
For instance, periodic polymers with a few long blocks easily undergo 
micro--phase separation and for them the frustrated states are suppressed
at equilibrium and in kinetics. There is a view in the scientific community
that complex random sequences also permit more refined classification.
The identification of good folding sequences 
(see e.g. Ref. \cite{Irback}) is believed to be an
important prerequisite for unraveling the mysteries of proteins.

In the current paper the GSC method has been applied to a single chain
problem. However, since the kinetic equations (\ref{eq:kinmain}) 
are covariant it would
be relatively straightforward to apply it to solutions of many polymers.
Formation of nontrivial mesoscopic globules in copolymer solutions at
low concentrations has been predicted in the framework of the method
\cite{MesoTheor} and recently observed experimentally \cite{MesoExp}.
Generally, the GSC method may be viewed as extension to the realm of 
kinetics of the variational treatment, 
which is a better theory than standard mean--field
approaches due to a much larger number of variational parameters.
We should admit, however, that in the complete form numerical solution of the 
GSC equations
is computationally expensive for systems of a few hundred of monomers.
Nevertheless,
as we have recently shown \cite{MesoTheor}, in some approximation
the GSC equations can be reduced to those of a simple mean--field theory 
such as the  Flory--Huggins one, and some extra corrections to the latter.
Importantly,
the formalism in terms of the mean--squared distances is valid for description
of the extended coil as well as of the globular states, whereas the density
formalism theories are limited to relatively high densities of 
the system. 
The weakest point of the method is the Gaussian form of the trial Hamiltonian,
and thus of the correlation functions,
a matter that has been discussed at some length in the past
\cite{GscKinet,GscHomKin}.
However, based on the covariant form of the kinetic equations derived here,
we hope, in some not too distant future,
to take into account the non--Gaussian corrections 
in an analogy to the treatment of the hard--core repulsion for van
der Waals systems. 

The main strength of the revised GSC approach lies in the ability
to describe a given heteropolymer sequence of finite length, rather than an
{\it infinite ensemble}
of sequences characterised in a certain probabilistic way.
This is an inevitable preliminary step in developing
adequate techniques for theoretical modelling of complex biopolymers.
Understanding of
the relation between the chemical composition (primary sequence)
and the 3-d equilibrium (tetriary) structure, 
as well as of the kinetics of folding,
in proteins is one of the great challenges of the modern biotechnological
science. 
We hope that methods like the one we have presented here will take their
right place in the collection of new tools for bioinformatics.

\acknowledgments

The authors acknowledge interesting discussions with 
Professor C.A.~Angell, Professor K.~Kawasaki, Professor G.~Parisi,   
and our  colleagues Dr A.V.~Gorelov and A.~Moskalenko. 

\appendix

\section{Inclusion of hydrodynamics}\label{app:hydro}

The Langevin equation with account of the hydrodynamic interaction
includes the Oseen tensor \cite{Edwards},
\begin{equation}
\frac{d}{dt}X_n^\alpha = \sum_{\alpha',n'}H^{\alpha \alpha'}_{nn'}[X(t)]\, 
\phi^{\alpha'}_{n'},
\end{equation}
where $\phi^\alpha_n$ is the right--hand side of Eq. (\ref{eq:Lang})
and the noise has the second momentum proportional to the inverse Oseen
tensor.

In the preaveraged approximation we have,
\begin{equation}
\langle H^{\alpha\alpha'}_{mm'} \rangle = \delta^{\alpha\alpha'} \xi_{mm'}, 
\quad
\xi_{mm'}= \frac{\delta_{mm'}}{\zeta_b}+\frac{1-\delta_{mm'}}
{ 3(2\pi^3)^{1/2}\eta_s\,D_{mm'}^{1/2}},
\end{equation}
and the analogs of formulae (A5,A6) 
of Ref. \cite{GscBlock} will be,
\begin{eqnarray}
\frac{1}{2} \frac{d}{dt} D_{nn'} &=& \sum_{m'}(\xi_{nm'}-\xi_{n'm'})
(\Gamma_{nm'}-\Gamma_{n'm'}), \\
\Gamma_{nn'} &=& -\frac{2}{3}\sum_{m''}D_{nm''}\frac{\partial
{\cal A}}{\partial D_{n'm''}}= k_B T \delta_{nn'} + \sum_{m''} D_{nm''}
V_{n'm''}.
\end{eqnarray}
Explicit expressions for a few first terms in 
the virial expansion of the effective potentials,
\begin{equation}
\label{eq:V}
V_{mm'} = -\frac{2}{3}\frac{\partial {\cal E}}{\partial D_{mm'}} 
\end{equation}
may be found in Appendix 
\ref{app:der}.

\section{The variational principle}\label{app:var}

We introduce the trace as the integration over
all monomers coordinates subject to the constraint that
the reference point is fixed in the centre-of-mass of the chain,
\begin{equation} \label{eq:Tr}
{\rm Tr} \equiv \int \prod_{n=0}^{N-1}d{\bf X}_n\,\delta(\sum_n {\bf X}_n).
\end{equation}
Then the partition function is obtained as 
$Z_{tot}={\rm Tr} \exp(-H/k_B T)$ with $H$
given by Eq. (\ref{eq:H}). 
The delta--function in (\ref{eq:Tr}) removes the trivial divergence
in $Z_{tot}$ due to the translational invariance.

We choose the trial Hamiltonian as
an arbitrary linear combination,
\begin{equation}
\frac{H_0}{k_B T} = \frac{1}{2}\sum_{mm'} K_{mm'} {\bf X}_m {\bf X}_{m'}+
\sum_m {\bf J}_m {\bf X}_m,
\end{equation}
where we have also introduced arbitrary sources ${\bf J}_m$.
Using the delta--function one can exclude, say, ${\bf X}_0$ and derive,
\begin{equation} \label{Z0}
Z_0[J]=(2\pi)^{3(N-1)/2}({\rm det}\tilde{K}^{(N-1)})^{-3/2}
\exp\left(\frac{1}{2}\sum_{mm'}{}'({\bf J}_m-{\bf J}_0) 
(\tilde{K}^{-1})_{mm'}({\bf J}_{m'}-{\bf J}_0)
\right),
\end{equation}
where the $N-1$ dimensional matrix $\tilde{K}$ is
\begin{equation}
(\tilde{K})_{nn'}=K_{nn'}+K_{00}-K_{n0}-K_{0n'}.
\end{equation}
We can calculate the averages by,
\begin{equation}
{\cal F}_{nn'} = \frac{1}{3}\langle {\bf X}_n {\bf X}_{n'} \rangle_0 =
\frac{1}{3}\frac{\partial^2 
\log Z_0[J]}{\partial {\bf J}_n \partial {\bf J}_{n'}}=(\tilde{K}^{-1})_{nn'},
\end{equation}
where again the latter identity holds only for $n,n'\not= 0$. From 
these quantities it is straightforward to obtain the 
mean squared distances using Eq. (\ref{eq:Ddef})
which also holds only for $n,n'\not= 0$.

Finally, we want to express all independent parameters of the
matrix $\tilde{K}$ via the quantities $D_{mm'}$.
For this we compute the following sums applying the above delta--function
constraint,
\begin{equation}
\sum_{mm'} D_{mm'} = 2N\sum_m {\cal F}_{mm}, \quad
\sum_m D_{mn} = \frac{1}{2N}\sum_{mm'} D_{mm'}+N {\cal F}_{nn}.
\end{equation}
Substituting ${\cal F}_{nn}$ from the second formula here to Eq. (\ref{RtoDD})
and recalling the relation (\ref{D4})
we derive the desired inverse relation (\ref{RtoDD}) for
$(\tilde{K}^{-1})_{nn'}={\cal F}_{nn'}$.

\par From the partition function (\ref{Z0}) we obtain the ``entropic'' term
${\cal A}_0 \equiv -T{\cal S}=-k_B T\, \log Z_0[0]$ that yields 
precisely Eq. (\ref{Sr}). 
The Gibbs--Bogoliubov variational principle is then based
on minimising this variational free energy ${\cal A}
={\cal A}_0+\langle H-H_0\rangle_0$ with respect to the $N(N-1)/2$
independent variational parameters $D_{mm'}$.

\subsection{Entropy of the homopolymer}

For the homopolymer we have the translational invariance along
the chain so that $D_{nn'}=D_k$, where $k=|n-n'|$.
Then the matrix (\ref{RtoDD}) may be rewritten as,
\begin{equation}
R_{nn'}=R_g^2-\frac{1}{2}D_{nn'}, \quad R_g^2=\frac{1}{2N}\sum_k D_k.
\end{equation}
Let us apply a unitary transformation to the Fourier coordinates
generated by the matrix,
\begin{eqnarray}
f^{q}_n=\frac{1}{N^{1/2}} \exp\left(-\frac{2\pi i qn}{N}\right).
\end{eqnarray}
By a direct evaluation and recalling the relation for the normal
modes,
 \begin{equation}
-\frac{1}{2N^{1/2}}\sum_k {\rm Re}\, f^q_k\, D_k = {\cal F}_q,
\end{equation}
we see that the matrix $\tilde{R}=f^{\dagger}\bullet R \bullet f$
is diagonal with the eigenvalues $\tilde{R}_{00}=0$ and
$\tilde{R}_{qq}=N{\cal F}_q$.
Thus, the logarithm of determinant of the $(N-1)$-dimensional 
submatrix $\det' \tilde{R}$ yields the standard ``entropy'' of
the homopolymer (see Eq. (3) in Ref. \cite{GscTor}) up to a trivial constant.

\section{Formulae for the effective potentials}
\label{app:der}

Two first terms in the virial expansion of 
the derivatives (\ref{eq:V}) of the mean energy are 
explicitly given by,
\begin{eqnarray}
V^{(2)}_{mm'} &=& (1-\delta_{mm'})\frac{\hat{u}^{(2)}_{mm'}}
{D^{5/2}_{mm'}}-\delta_{mm'}\sum_{\stackrel{n}{n\not=m}}
\frac{\hat{u}^{(2)}_{mn}}{D^{5/2}_{mn}}, \label{eq:c1} \\
V^{(3)}_{mm'} &=& (1-\delta_{mm'})3\hat{u}^{(3)}
\sum_{\stackrel{n}{n\not=m,m'}} \frac{D_{mn,m'n}}
{(\det\Delta^{(2)})_{mm'n}^{5/2}}
-\delta_{mm'}\frac{3}{2}\hat{u}^{(3)}\sum_{\stackrel{nn'}{n\not= n'\not= m}}
\frac{D_{nn'}}{(\det\Delta^{(2)})_{mnn'}^{5/2}} , \label{eq:c2}
\end{eqnarray}
where $(\det\Delta^{(2)})_{mm'm''}=D_{mm'}D_{m''m'}
-D_{mm',m''m'}^2$.

It is usually assumed that
for negative $\bar{u}^{(2)}$ the stability of the system
is ensured by the positiveness of $u^{(3)}$. 
It turns out however that for the whole range of $u^{(2)}_{mm'}$ this
does not hold and there are additional pathological solutions with singular 
free energy. The reason for this deficiency of the model (\ref{eq:H})
is that we have discarded the terms with two coinciding indices in
the three--body contribution. This standard trick is not satisfactory therefore,
but fortunately it could be remedied by using another prescription for
these terms:
\begin{eqnarray} \label{eq:c3}
{\cal E}_3&=& u^{(3)}\sum_{m\not=m'\not=m''}\biggl\langle
\delta({\bf X}_{m}-{\bf X}_{m'}) \, \delta({\bf X}_{m''}-{\bf X}_{m'})
\biggr\rangle
+ 3 u^{(3)}\sum_{m\not=m'}\biggl\langle
\delta({\bf X}_{m}-{\bf X}_{m'}) \biggr\rangle^2.
\end{eqnarray}
In Ref. \cite{GscC3Unp} we show that the addition of the latter
term, which is subdominant in the large $N$ limit anyway, 
does not change the earlier results for the homopolymer,
and even quantitatively improves the agreement with known results
for the dense globular state \cite{GrosbKhokh}. 
Importantly, for heteropolymers this
term removes spurious solutions and makes the theory well defined. 
The corresponding contributions to the mean energy and the effective
potentials are,
\begin{equation} \label{eq:c4}
{\cal E}^{(3')}= 3\hat{u}^{(3)}\sum_{m\not=m'} D_{mm'}^{-3}, \qquad
V^{(3')} = (1-\delta_{mm'}) 6\hat{u}^{(3)}D^{-4}_{mm'}
-\delta_{mm'}6\hat{u}^{(3)}\sum_{\stackrel{n}{n\not=m}} D^{-4}_{mn}.
\end{equation}


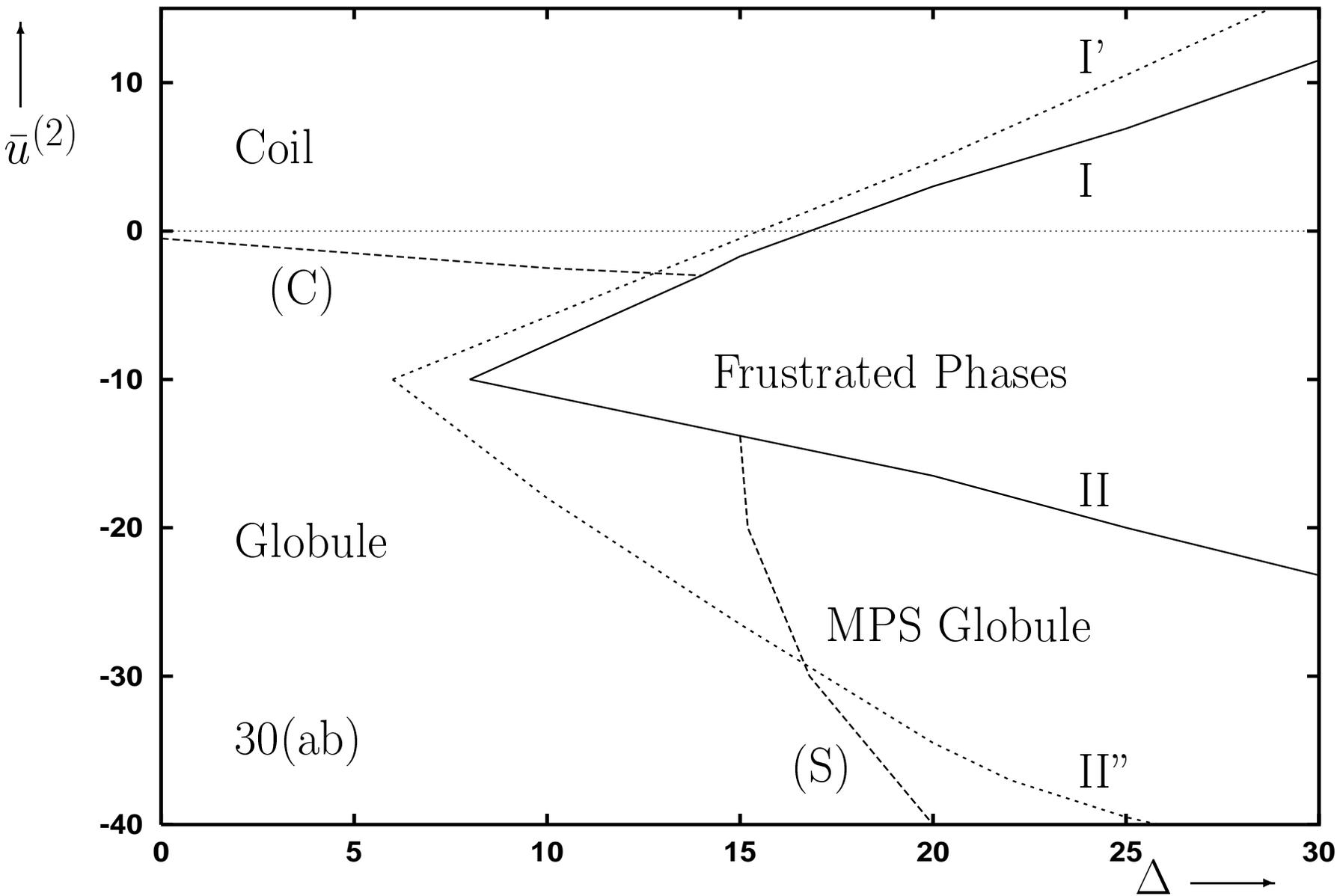
\begin{figure}
\caption{ 
The phase diagram of copolymer sequence 30(ab) (``short'' blocks) in terms
of the mean second virial coefficient, $\bar{u}^{(2)}$, and the amphiphilicity,
$\Delta$ (both in units $k_B T {\cal L}^3$). 
Curves (C) and (S) correspond respectively to the collapse and the
micro--phase separation continuous transitions. Curves (I) and
(II) correspond to discontinuous transitions to the
frustrated phases. ``Spinodal'' curves (I') and (II'') bound the regions
of metastability of the frustrated states. Transition curves and boundaries 
distinguishing different frustrated sates are not depicted.
}\label{fig:1}
\end{figure}

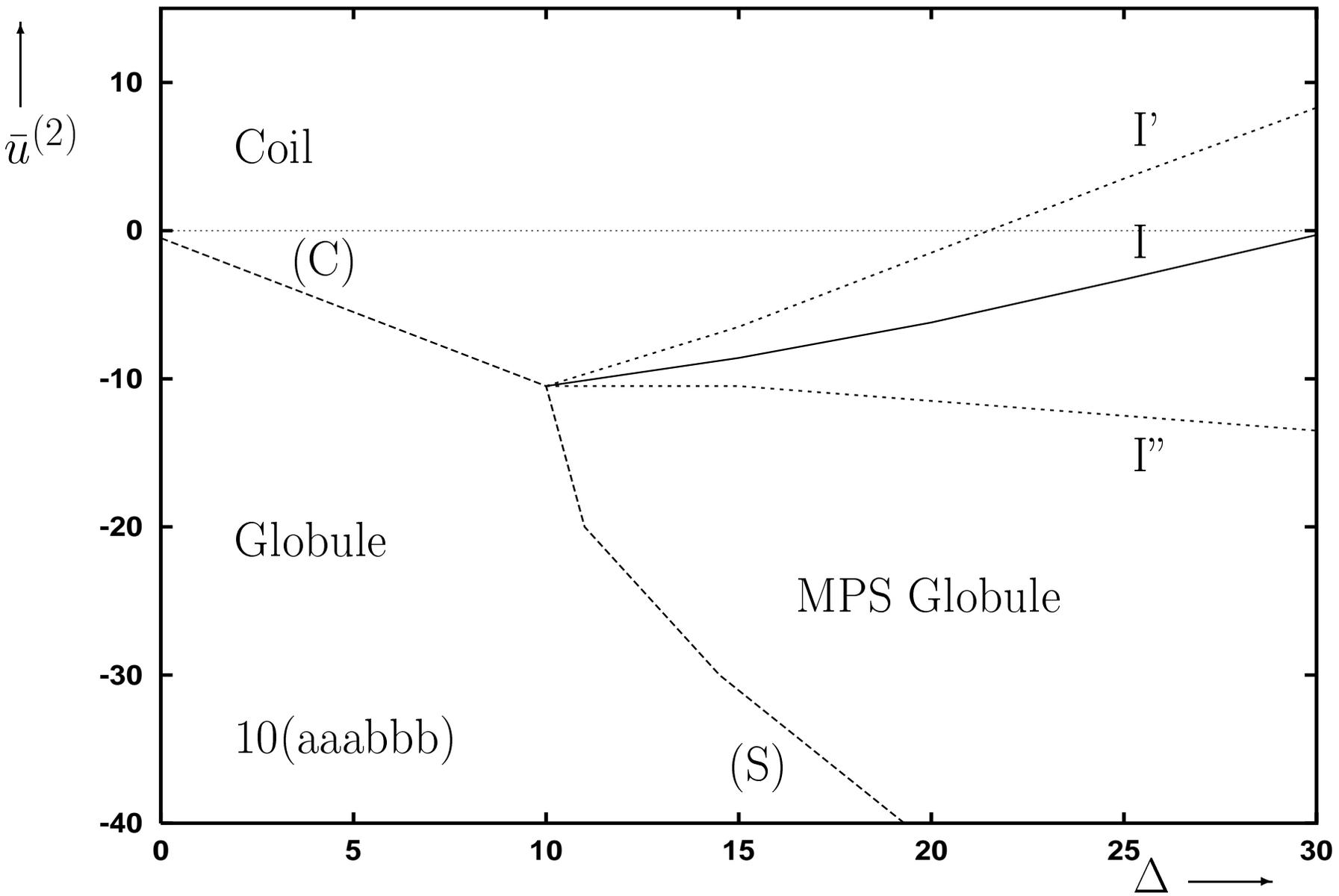
\begin{figure}
\caption{ 
The phase diagram of copolymer sequence 10(aaabbb) (``long'' blocks) in terms
of the mean second virial coefficient, $\bar{u}^{(2)}$, and the amphiphilicity,
$\Delta$ (both in units $k_B T {\cal L}^3$). 
For large values of $\Delta$ the collapse transition becomes 
discontinuous (curve (I)) and it is accompanied by micro--phase separation
(see also Figs.~\ref{fig:6}-\ref{fig:7}). 
Curves (I') and (I'') are spinodals.
}\label{fig:2}
\end{figure}

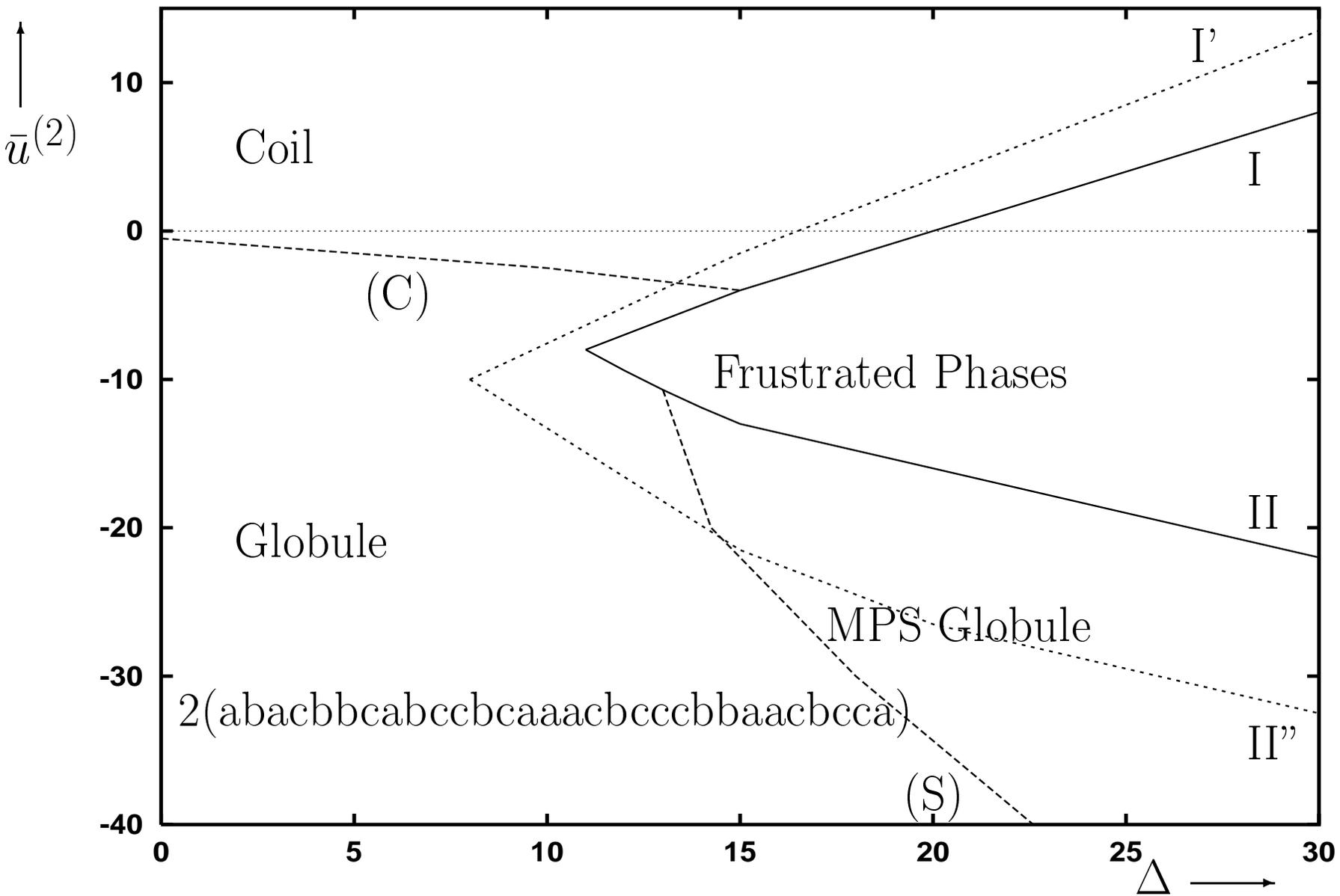
\begin{figure}
\caption{ 
The phase diagram of the ``random'' sequence 
2(abacbbcabccbcaaacbcccbbaacbcca) in terms
of the mean second virial coefficient, $\bar{u}^{(2)}$, and the amphiphilicity,
$\Delta$ (both in units $k_B T {\cal L}^3$). 
Other notations are as in Fig.~\ref{fig:1}.
}\label{fig:3}
\end{figure}

\begin{figure}
\caption{ 
Diagrams of the mean squared distances matrix $D_{mm'}$ for 
the ``short'' blocks copolymer 30(ab)
at $\Delta = 20$. Diagrams (a-d) correspond
respectively to $u^{(2)} = 15$, $-21$, $-30$ and $-40$
(in units $k_B T {\cal L}^3$).
Indices $m$, $m'$ start counting from the upper
left corner. Each matrix element, $D_{mm'}$ is denoted by a quadratic
cell with varying degree of black colour, the darkest and the
lightest cells corresponding respectively to the smallest and to the largest
mean squared distances. The diagonal elements are not painted since
$D_{mm} = 0$.
For the coil (Fig.~a) $D_{mm'}$ elements increase
monotonically on moving away from the diagonal towards 
half--ring distance along the chain.
In frustrated states (Figs.~b,c) $D_{mm'}$
possesses some number of clusters with monomers having smaller
distances between each other.
For the MPS globule (Fig.~d) $D_{mm'}$ reflects the structure of the
two--body interaction matrix $u^{(2)}_{mm'}$ and consists of similar elementary
cells.
}\label{fig:dmneq}
\end{figure}

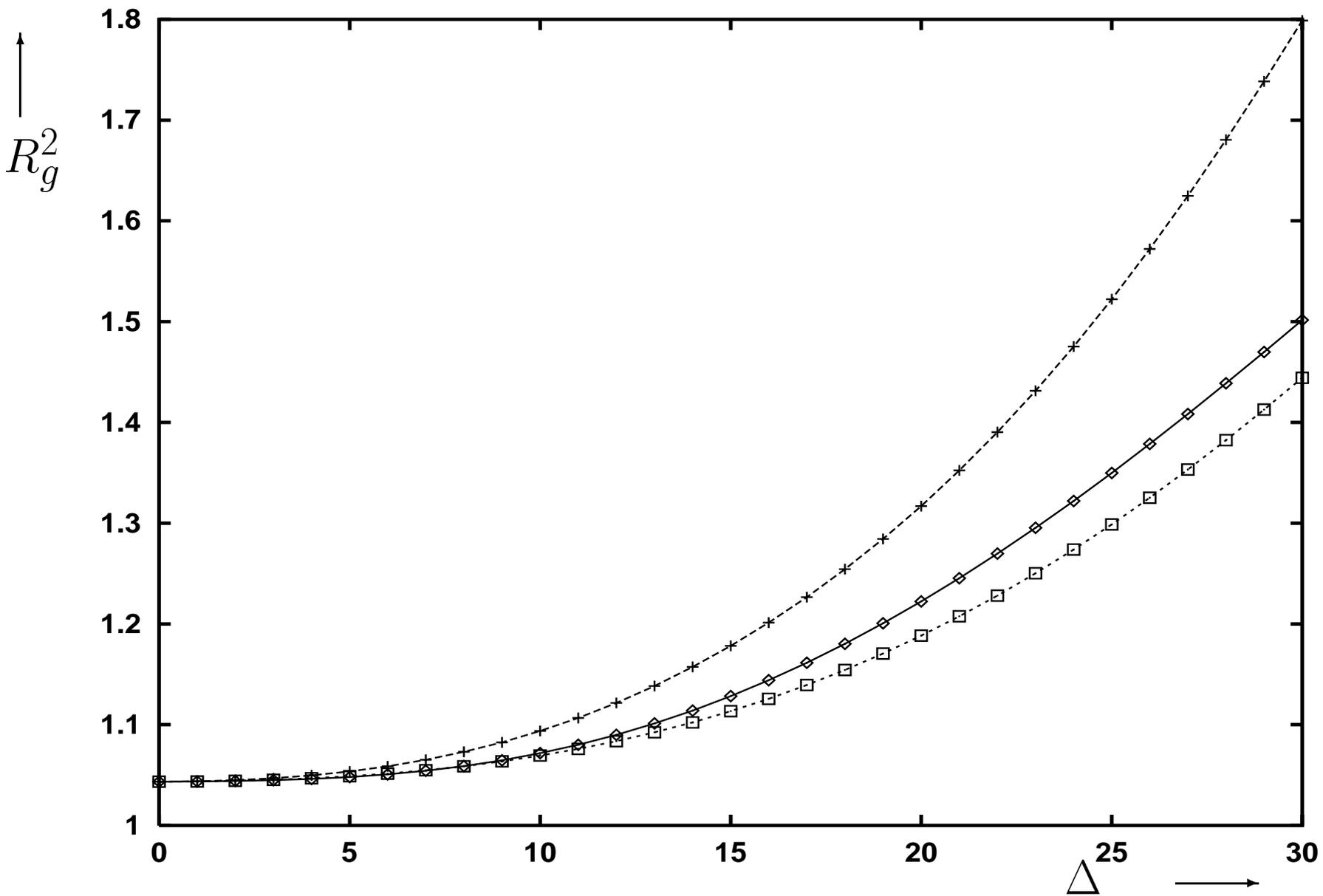
\begin{figure}
\caption{ 
Plot of the mean squared radius of gyration, $R_g^2$
(in units ${\cal L}^2$), vs the amphiphilicity,
$\Delta$ (in units $k_B T {\cal L}^3$), 
for different sequences (from top to bottom): 
``long'' blocks, ``short'' blocks and the
``random'' sequence. Here we have
fixed $\bar{u}^{(2)} = -40$.
}\label{fig:4}
\end{figure}

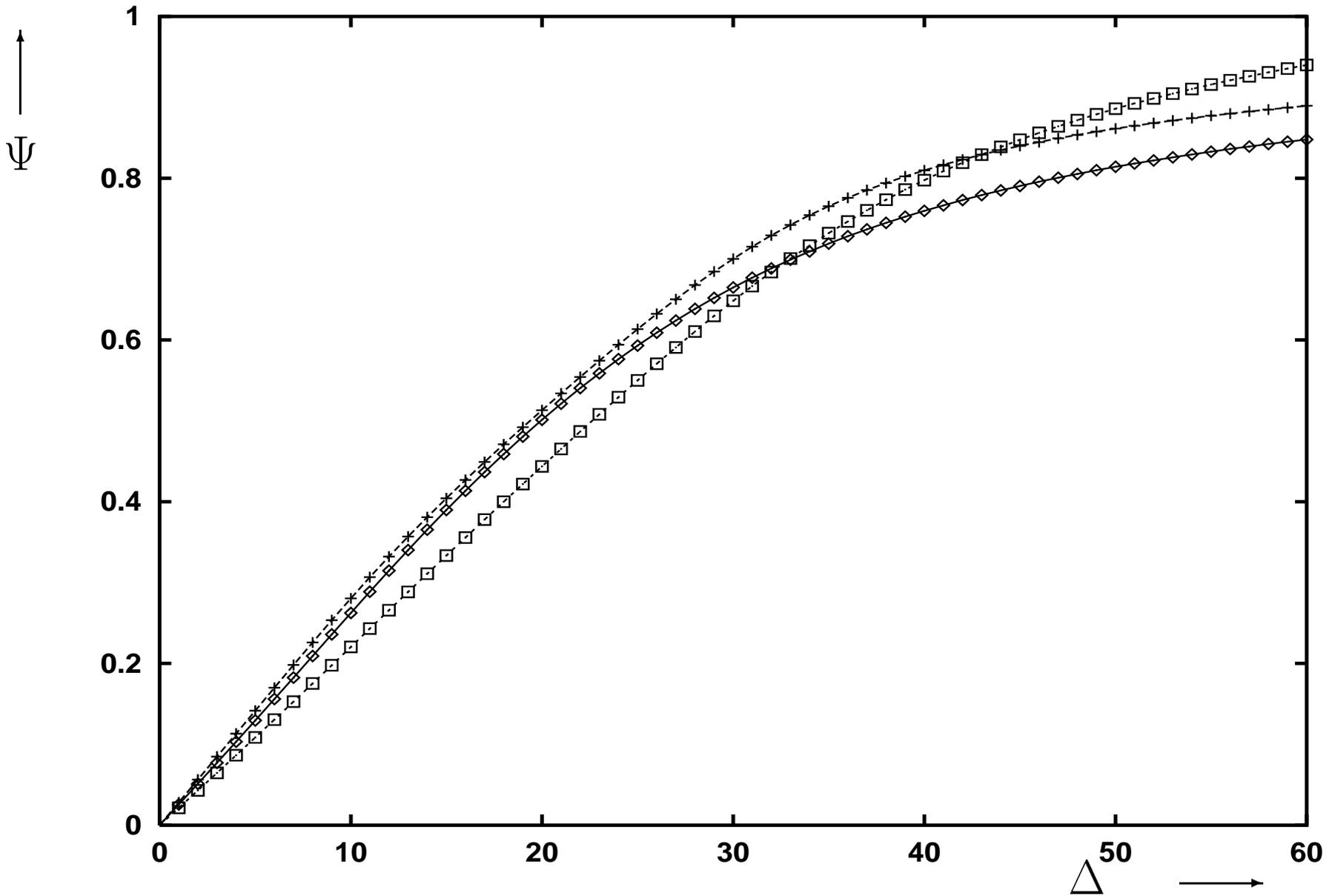
\begin{figure}
\caption{ 
Plot of the parameter of micro--phase separation, $\Psi$, vs the 
amphiphilicity,
$\Delta$ (in units $k_B T {\cal L}^3$), for different sequences:
 ``long'' blocks (pluses), ``short'' blocks (diamonds) and the
``random'' sequence (quadrangles). Here we have
fixed $\bar{u}^{(2)} = -40$.
}\label{fig:5}
\end{figure}

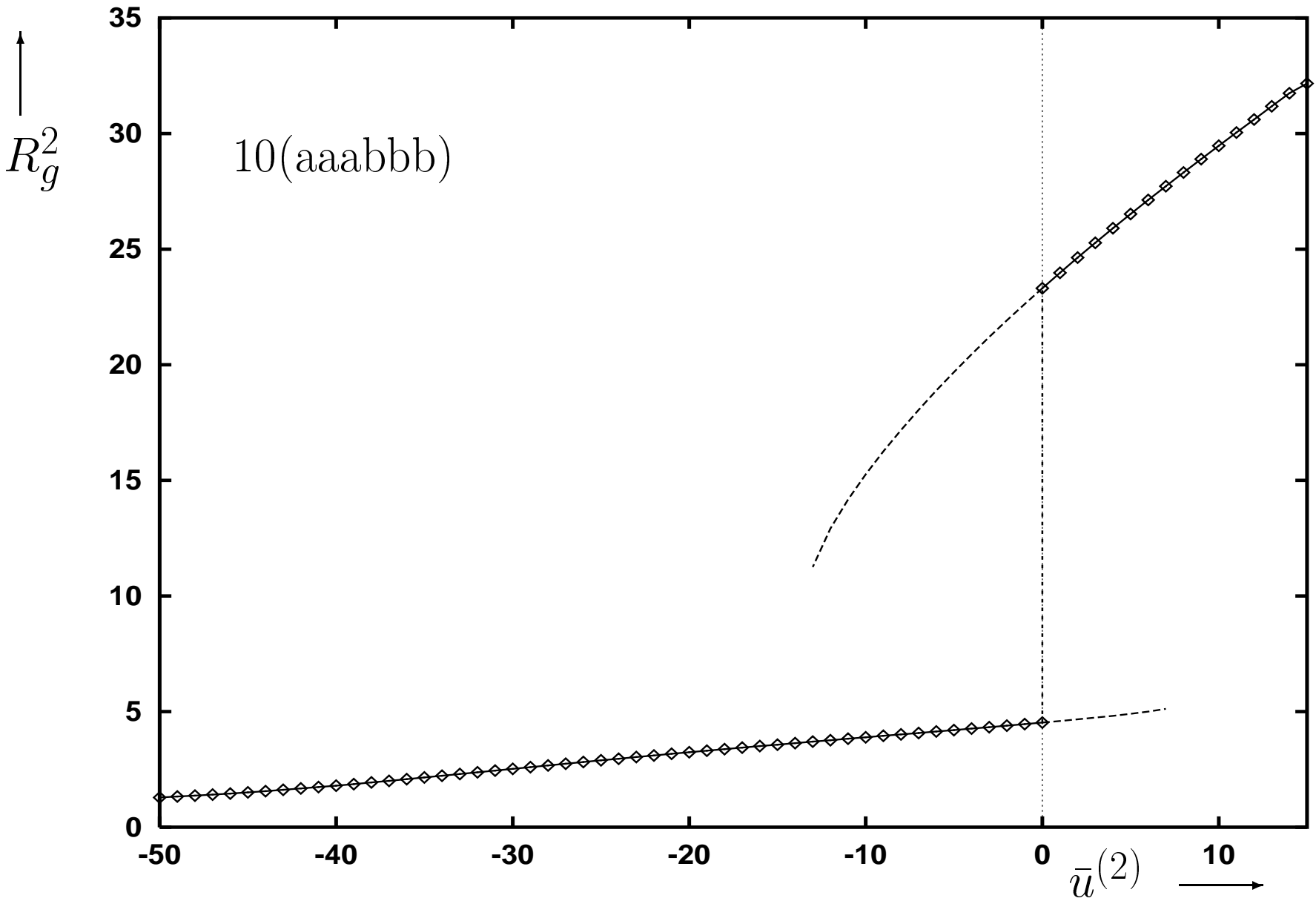
\begin{figure}
\caption{ 
Plot of the mean squared radius of gyration, $R_g^2$ 
(in units ${\cal L}^2$), vs the second virial
coefficient, $\bar{u}^{(2)}$ (in units $k_B T {\cal L}^3$),
for 10(aaabbb) copolymer. Here and in
Figs.~\ref{fig:7}-\ref{fig:9} 
$\Delta = 30$, solid lines correspond to values of observables in
the main free energy minimum and dashed lines --- in the metastable minima.
}\label{fig:6}
\end{figure}

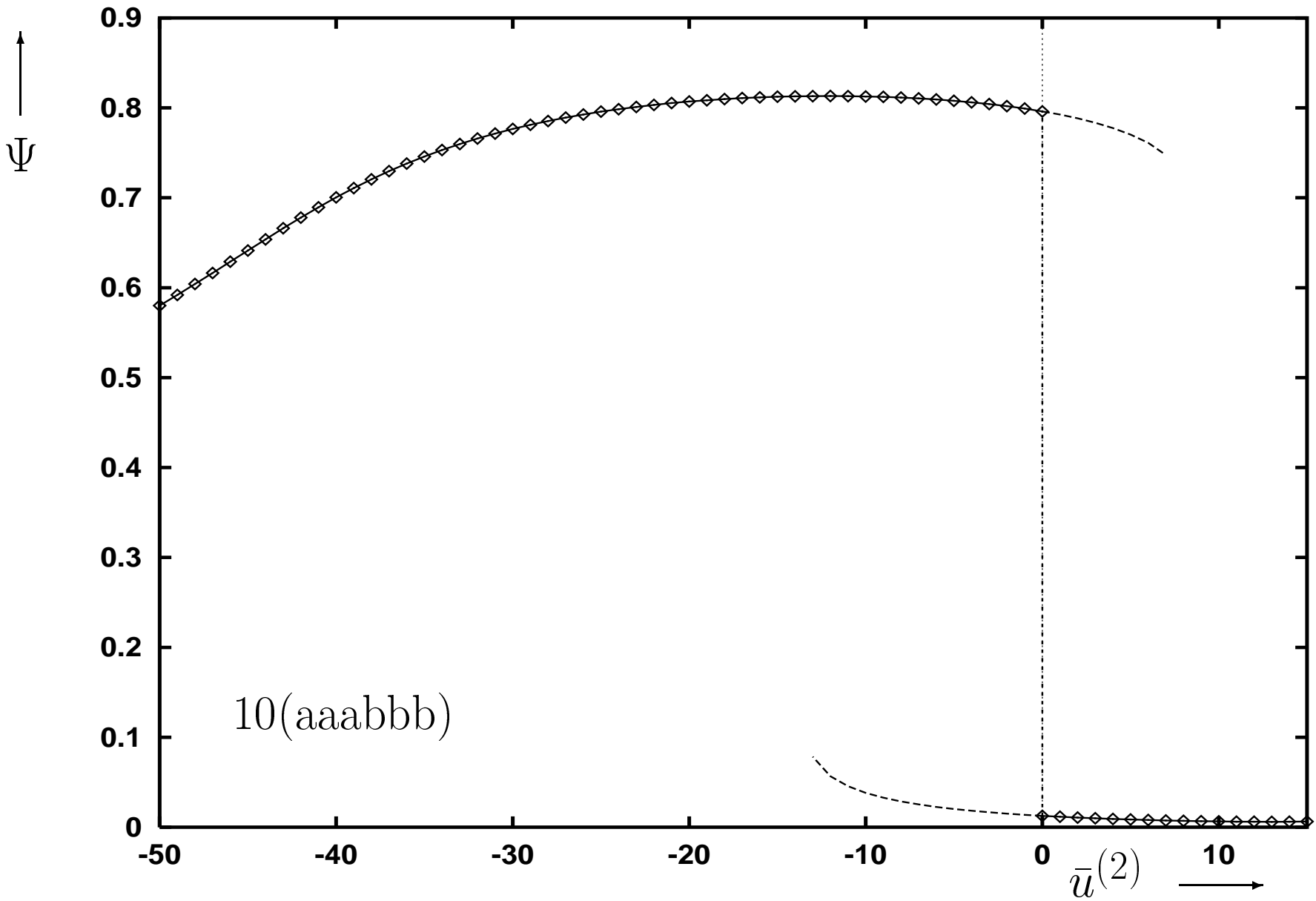
\begin{figure}
\caption{ 
Plot of the parameter of micro--phase separation, $\Psi$, vs the second virial
coefficient, $\bar{u}^{(2)}$ (in units $k_B T {\cal L}^3$), 
for 10(aaabbb) copolymer.
}\label{fig:7}
\end{figure}

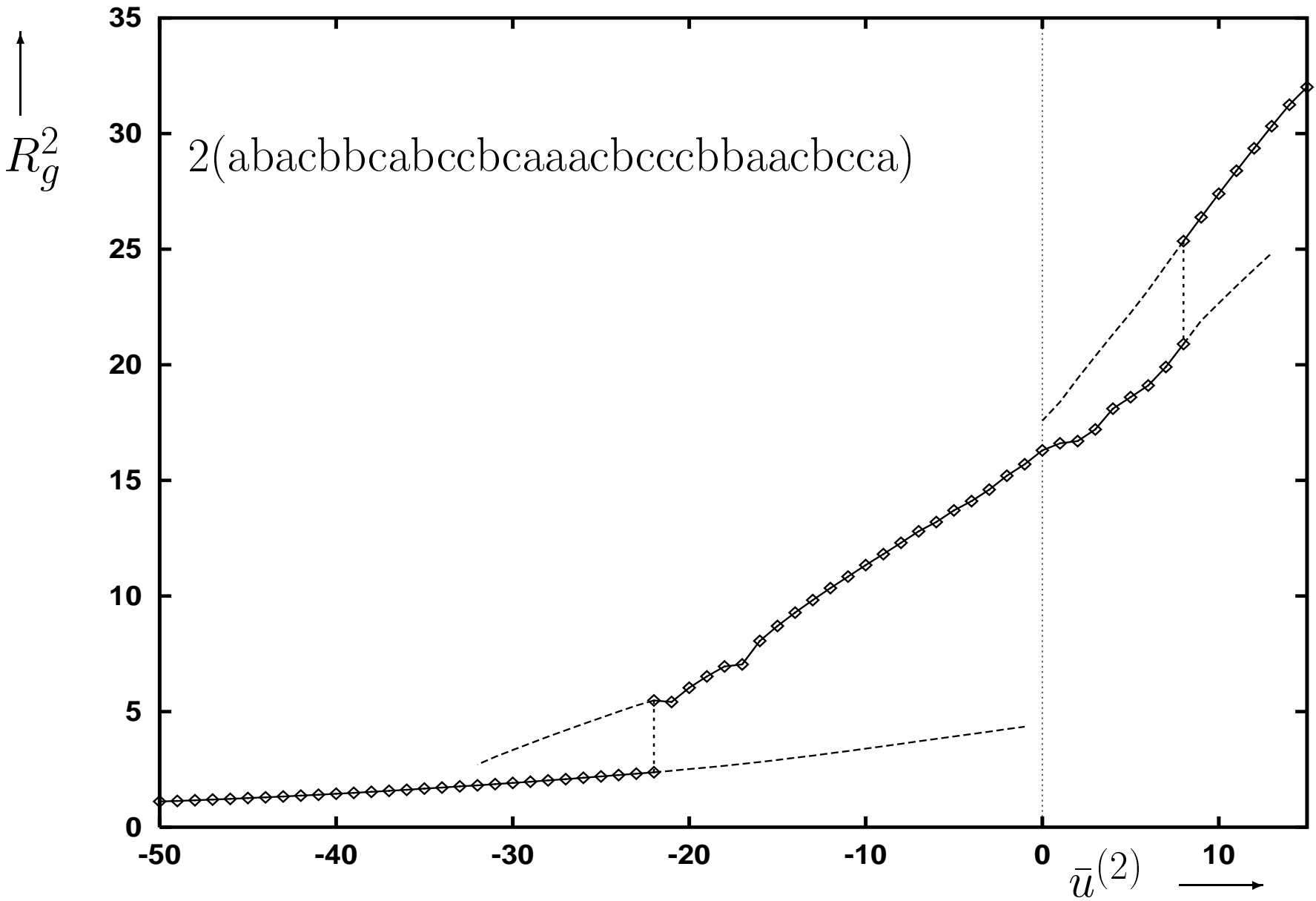
\begin{figure}
\caption{ 
Plot of the mean squared radius of gyration, $R_g^2$ (in units ${\cal L}^2$), 
vs the second virial
coefficient, $\bar{u}^{(2)}$ (in units $k_B T {\cal L}^3$),
for the ``random'' sequence.
}\label{fig:8}
\end{figure}

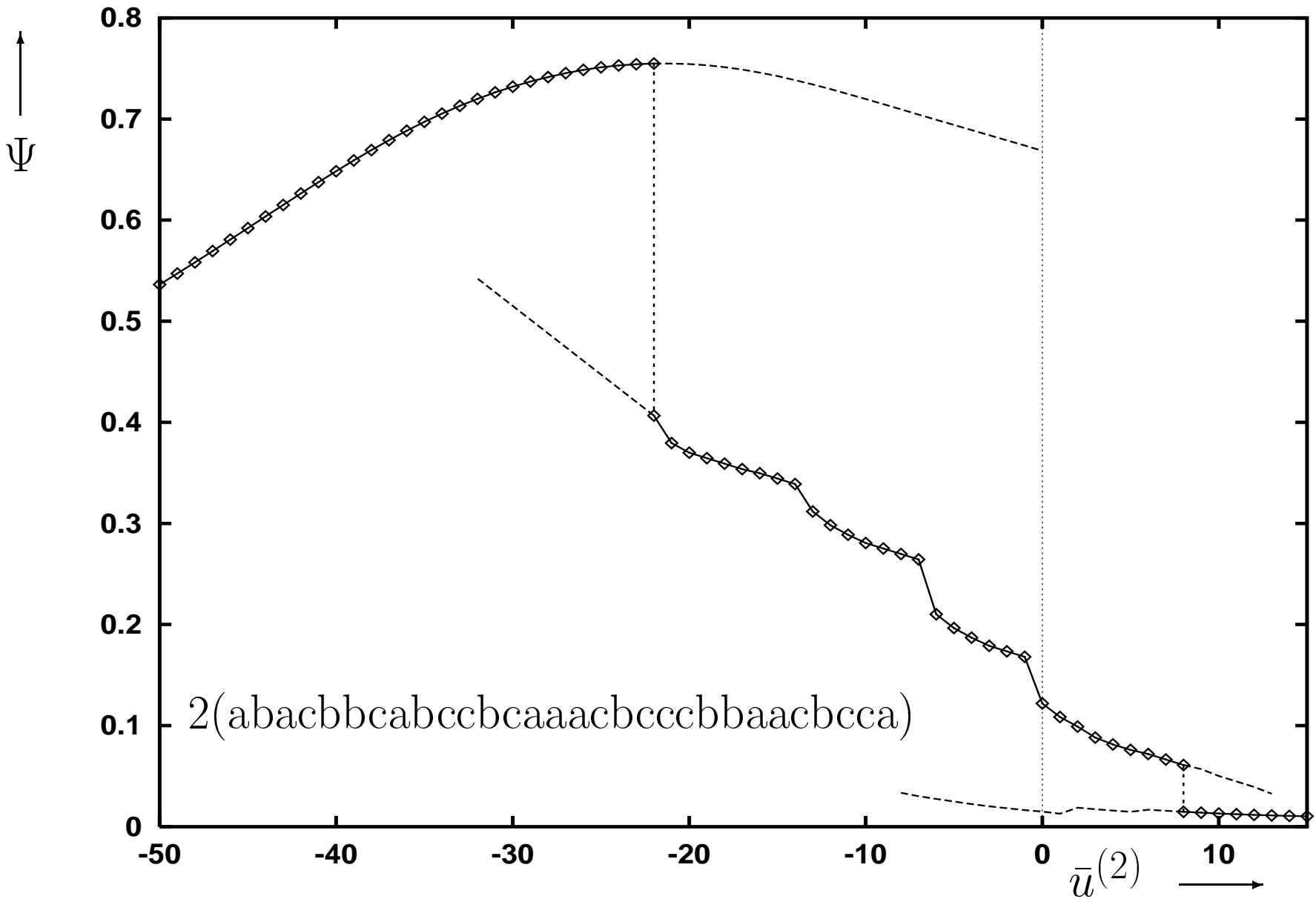
\begin{figure}
\caption{ 
Plot of the parameter of micro--phase separation, $\Psi$, vs the second virial
coefficient, $\bar{u}^{(2)}$ (in units $k_B T {\cal L}^3$),
for the ``random'' sequence.
}\label{fig:9}
\end{figure}

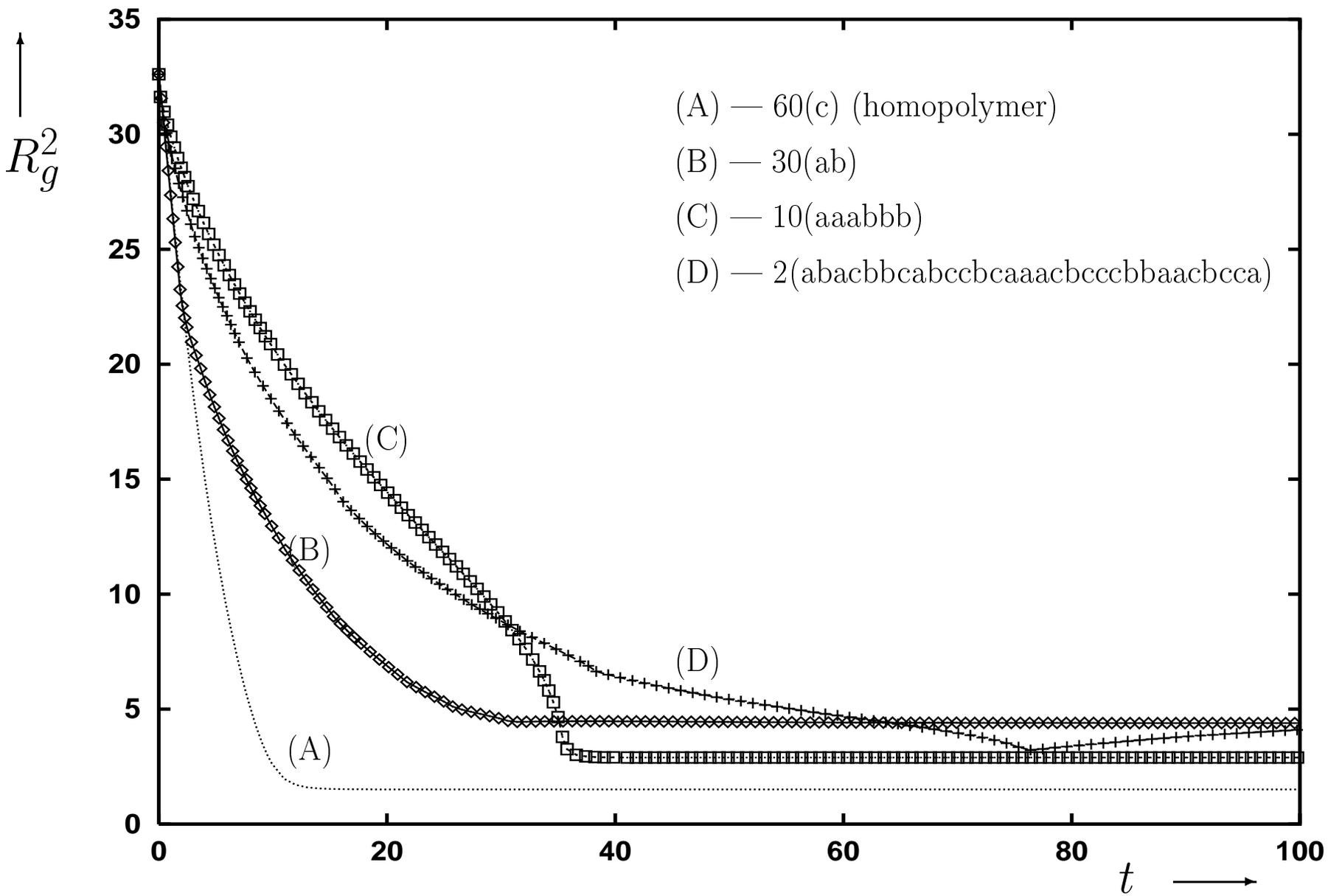
\begin{figure}
\caption{ 
Time evolution ($t$ is in units ${\cal T}$) 
of the mean squared radius of gyration, $R_g^2$ 
(in units ${\cal L}^2$), for different
sequences after an instantaneous quench from the coil state,
$\bar{u}^{(2)} = 15$ and $\Delta = 0$, to the region with 
$\bar{u}^{(2)} = -25$ and $\Delta = 30$.
}\label{fig:10}
\end{figure}

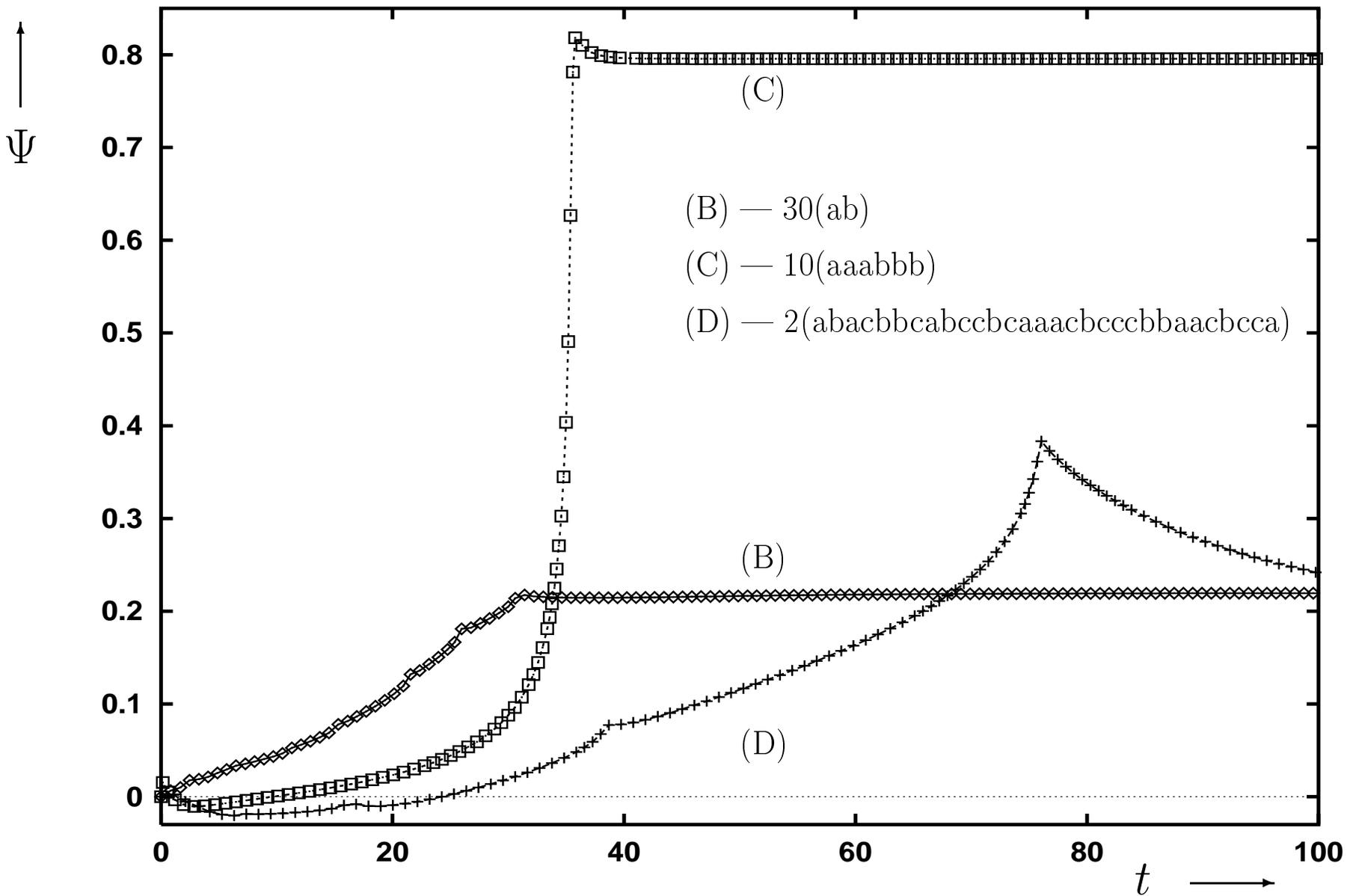
\begin{figure}
\caption{ 
Time evolution ($t$ is in units ${\cal T}$)
of the parameter of micro--phase separation, $\Psi$, 
for different sequences after the same quench as in Fig.~\ref{fig:10}.
}\label{fig:11}
\end{figure}

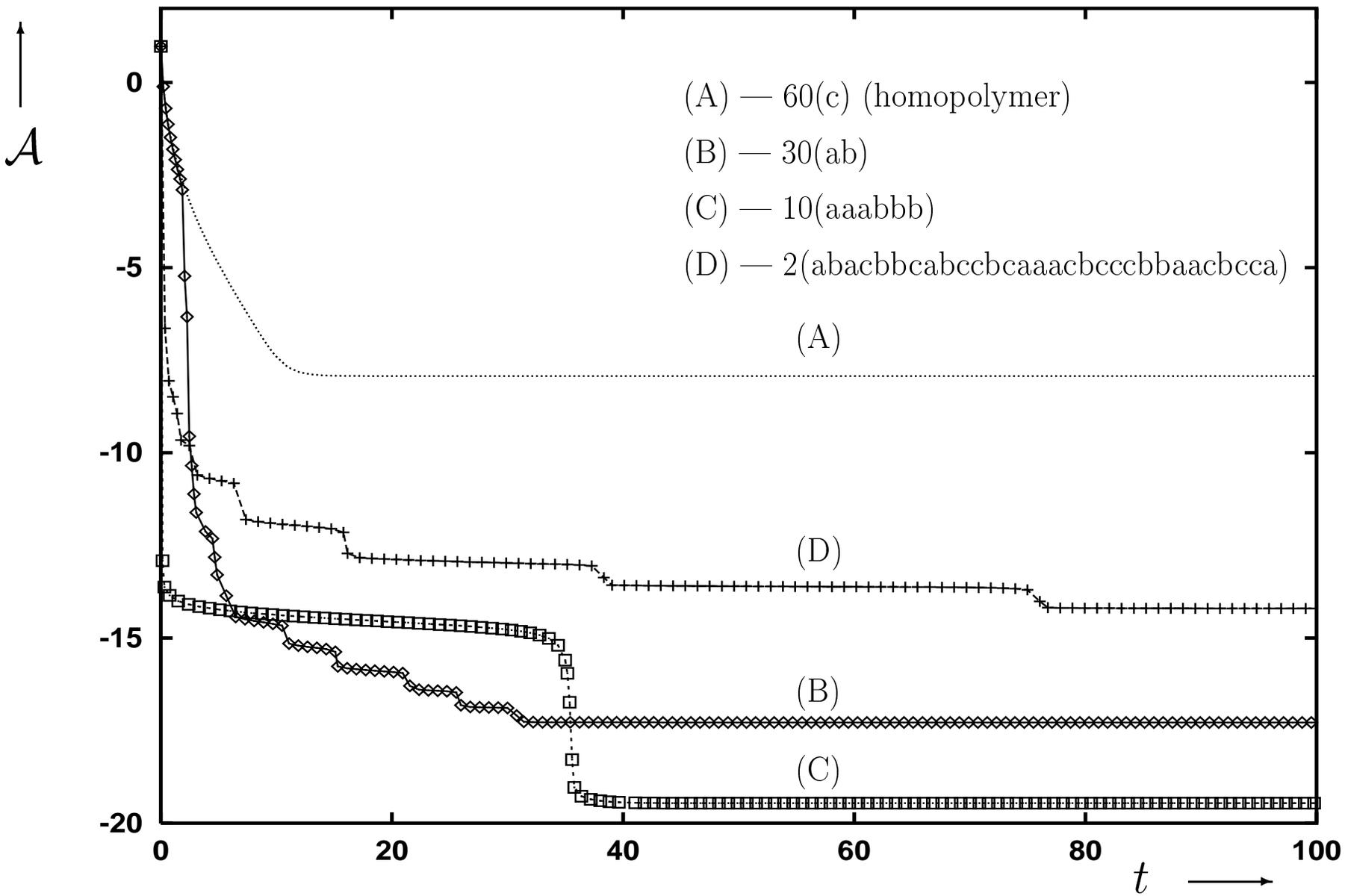
\begin{figure}
\caption{ 
Time evolution ($t$ is in units ${\cal T}$)
of the instantaneous free energy, ${\cal A}$ (in units $k_B T$),
for different sequences after the same quench as in Fig.~\ref{fig:10}.
}\label{fig:12}
\end{figure}

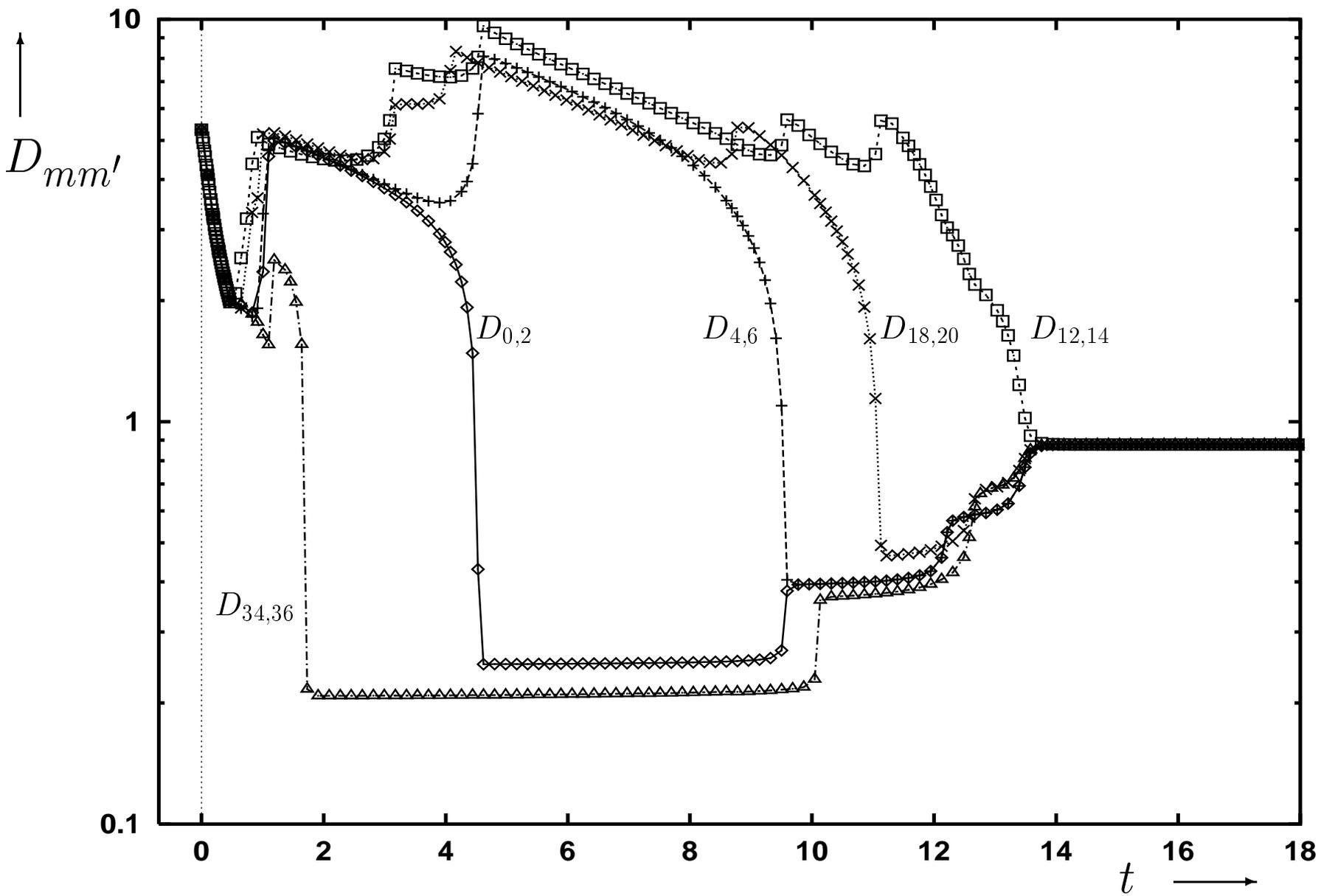
\begin{figure}
\caption{ 
Time evolution ($t$ is in units ${\cal T}$)
of the mean squared distances between nearest
`a' monomers $D_{2k,2k+2}(t)$ (in units ${\cal L}^2$)
for 30(ab) copolymer in
kinetics after the quench with the final two--body parameters: 
$\bar{u}^{(2)}=-50$ and $\Delta=30$.
}\label{fig:13}
\end{figure}

\begin{figure}
\caption{ 
Diagrams of $D_{mm'} (t)$ matrix for
the ``short'' blocks copolymer 30(ab) 
in kinetics after the same quench as in
Fig.~\ref{fig:13}. Diagrams (a-c) correspond respectively to the
following moments in time: $t = 4.6$, $11$, and $12.9$.
See also caption to Fig.~\ref{fig:dmneq} for more details.
The kinetic process proceeds
through formation of locally collapsed and phase--separated clusters.
The initial conformation is similar to Fig.~\ref{fig:dmneq}a, then 
some clusters appear, coalesce into larger ones,
until they eventually unify
forming the MPS globule (similar to Fig.~\ref{fig:dmneq}d).
}\label{fig:dmnkin}
\end{figure}


\begin{references}
\bibitem{Footnote1}
This is accomplished
by integrating out its degrees of freedom from the path integral
representation for the partition function \cite{GscRanPhd}.
\bibitem{Footnote2}
Note that one may use the symmetry properties $D_{mm'}=D_{m'm}$
and $D_{mm}=0$ only after the differentiation of the energy (\ref{eq:E}),
which is generally speaking expressed through the four--point 
correlation functions. One should use the standard differentiation rule,
$\partial D_{kk'}/\partial D_{mm'}\equiv \delta^k_m \delta^{k'}_{m'}$.
The structure of the kinetic equations is such that the symmetry properties
are automatically satisfied on the solution of 
these equations at each moment in time.
\bibitem{Footnote2b}
Apparently, for a finite length chain, the collapse transition is
characterised by a finite width, and therefore its representation in
Figs.~\ref{fig:1}-\ref{fig:3}
is sufficiently schematic with a suitable criterion 
\cite{GscTor} adopted to draw a definite curve for it.
\bibitem{Footnote3}
The curve (S) in Figs.~\ref{fig:1}-\ref{fig:3} is defined
by the condition $\Psi=1/2$, i.e. half of maximal possible value of the MPS
parameter.
\end{references}
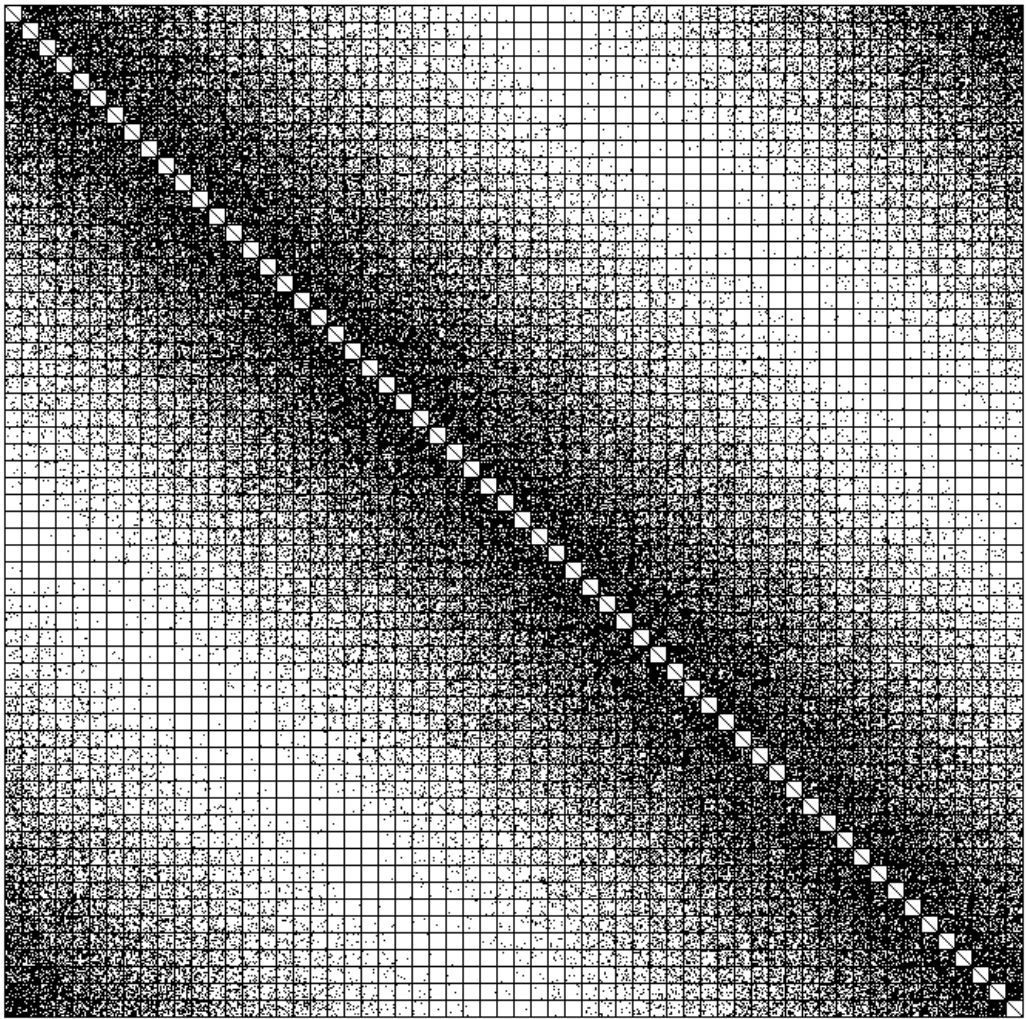
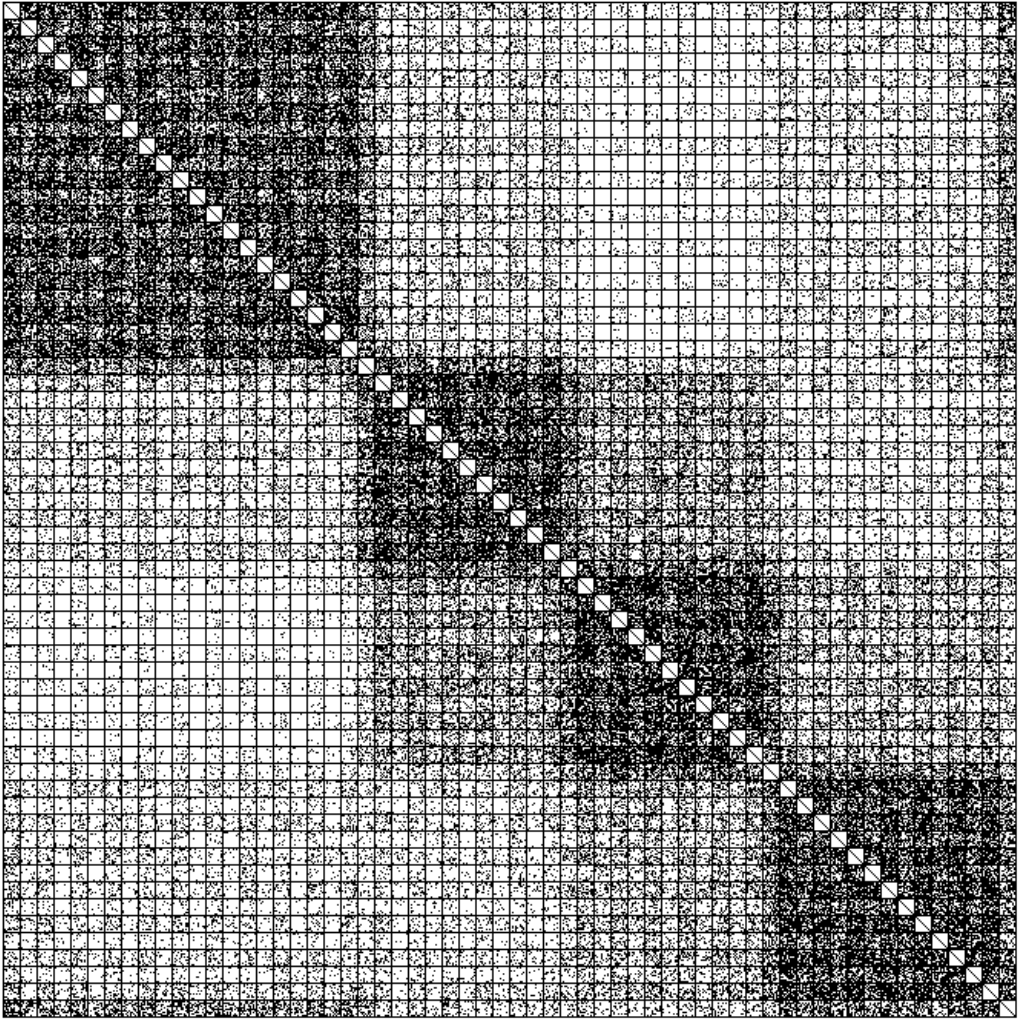
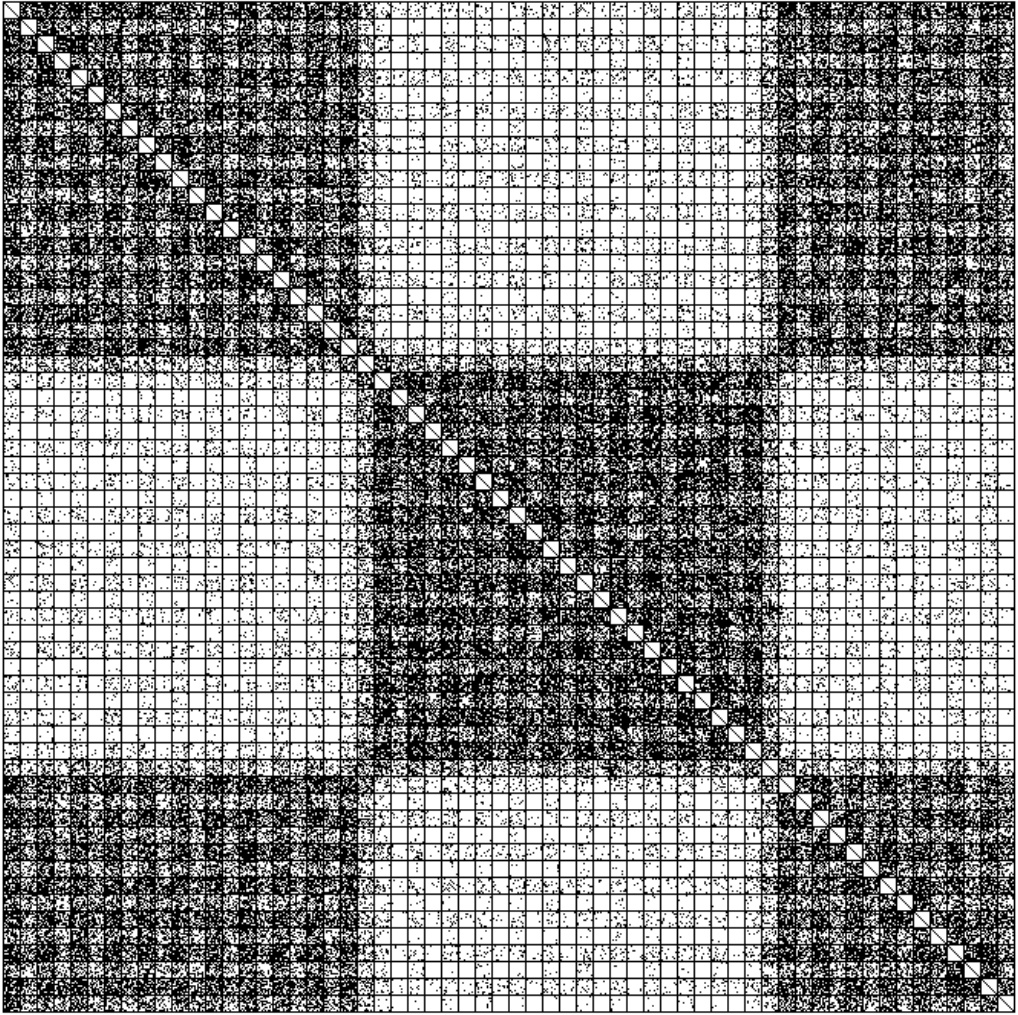
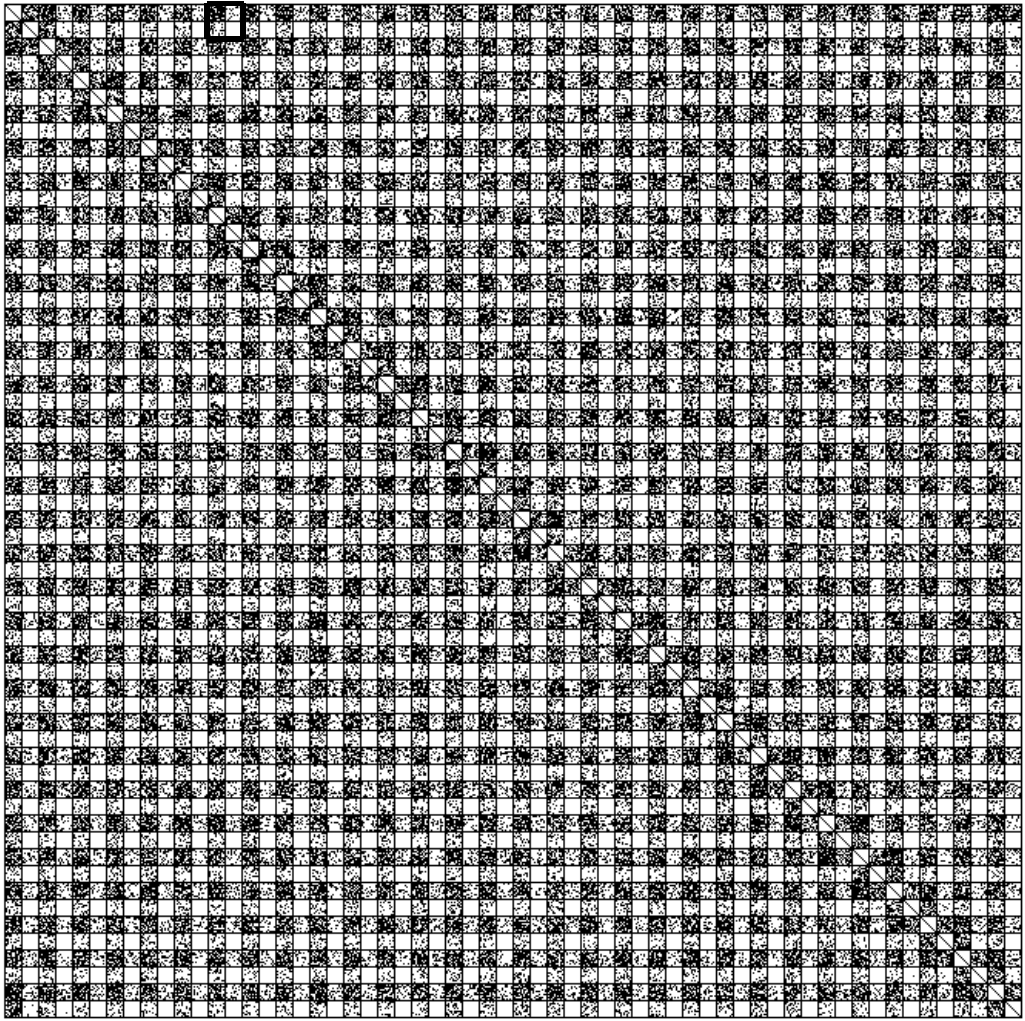
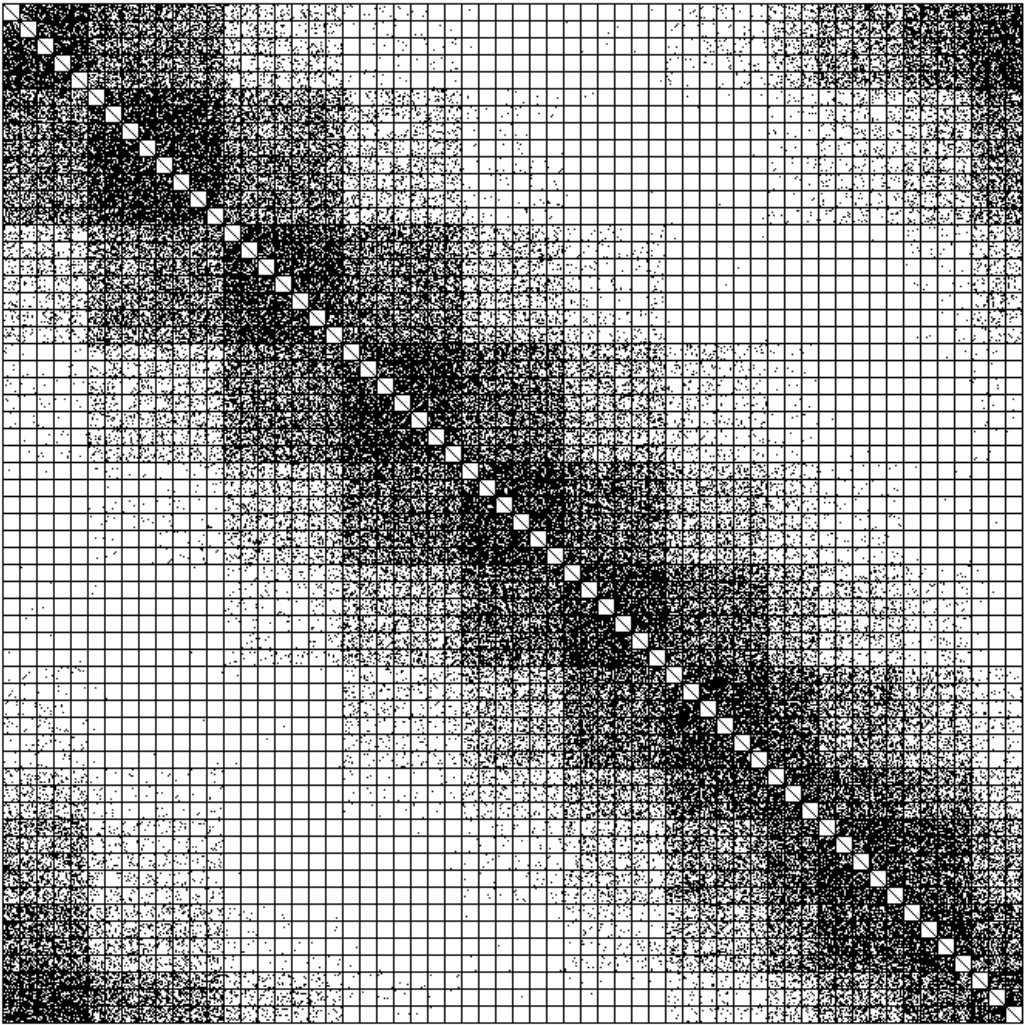
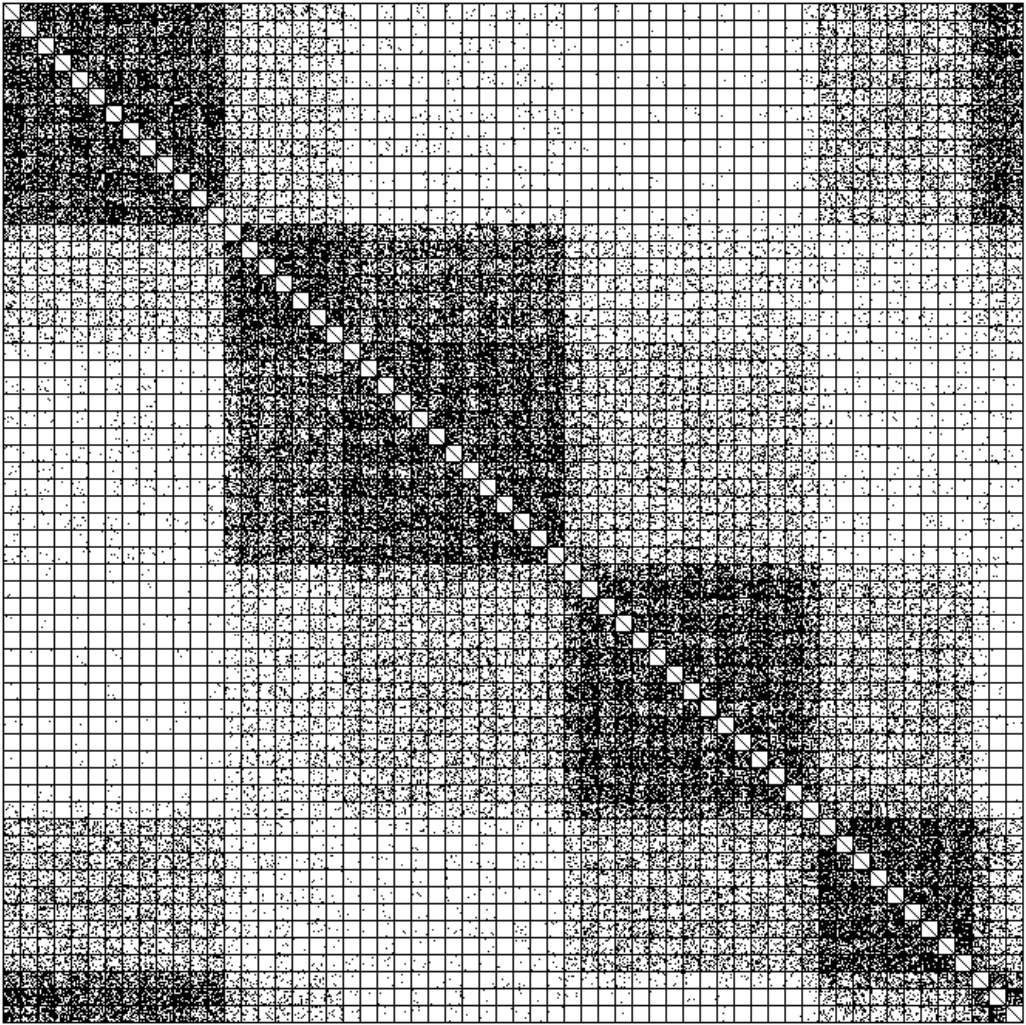
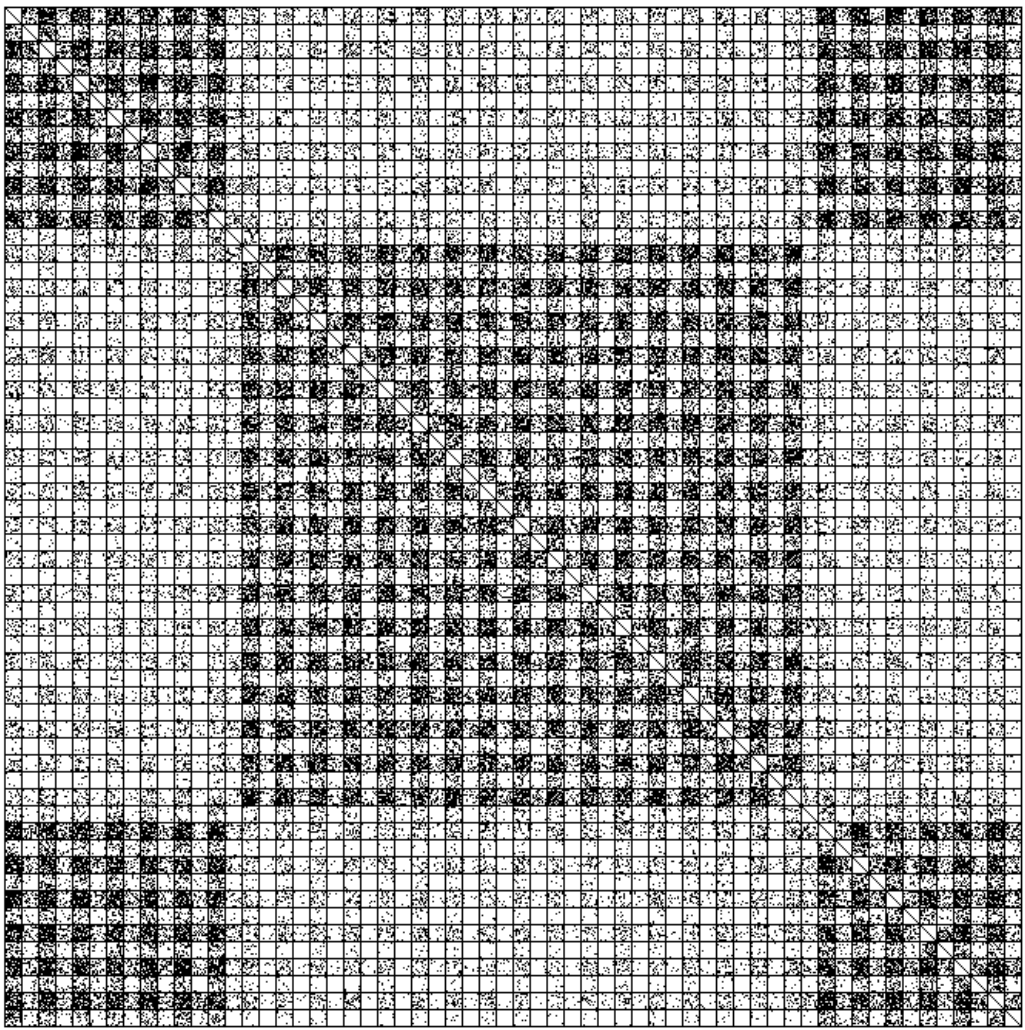
\end{document}